%% file: paper2.tex
\documentclass[onecolumn]{mn2e}  
\usepackage{epsf} 
\usepackage{amsmath} 
\begin{document}

\input{mymacros.tex}


\newcommand{\leftb}{<\!\!} \newcommand{\rightb}{\!\!>}

\newcommand{\oversim}[2]{\protect{\mbox{\lower0.5ex\vbox{%
  \baselineskip=0pt\lineskip=0.2ex
  \ialign{$\mathsurround=0pt #1\hfil##\hfil$\crcr#2\crcr\sim\crcr}}}}} 
\newcommand{\simgreat}{\mbox{$\,\mathrel{\mathpalette\oversim>}\,$}} 
\newcommand{\simless} {\mbox{$\,\mathrel{\mathpalette\oversim<}\,$}} 

\newcommand{\tcr}{t_{\rm cr}}

\newcommand{\erf}{{\rm erf}}

\newcommand{\sfe}{SFE} 

\newcommand{\df}{{\sc DF}}

\newcommand{\Fpk}{F_{\rm P/K}}

\def\emphasize#1{{\sl#1\/}}
\def\arg#1{{\it#1\/}}
\let\prog=\arg

\def\edcomment#1{\iffalse\marginpar{\raggedright\sl#1\/}\else\relax\fi}


\marginparwidth 1.25in
\marginparsep .125in
\marginparpush .25in

\reversemarginpar



%
\title[The impact of mass loss II. N-body integration]{The impact of mass loss on star cluster formation. II. 
Numerical N-body integration \& further applications} 
   \author[Boily \& Kroupa]{C.~M. Boily$^{1,\dagger}$ \&  P. Kroupa$^{2}$
\\          
$^{1}$Astronomisches Rechen-Institut, M\"onchhofstrasse 12-14 Heidelberg, D-69120 Germany 
\\
$^{2}$Institut f\"ur Theoretische Physik und Astrophysik der
Universit\"at Kiel, D-24098 Kiel, Germany 
\\ 
$^{\dagger}$Present address: Observatoire astronomique de Strasbourg, 11 rue de l'universit\'e, 67000 Strasbourg, France}

\maketitle

\begin{abstract} We subject  to 
an N-body numerical investigation our analysis of Paper I on  the survival of stellar clusters undergoing rapid mass 
loss. We compare  analytical tracks of bound 
mass-fraction {\it vs} star formation efficiency $\epsilon$ to those obtained 
with N-body integration.  We use these to argue that stellar clusters
must develop massive cores of high-binding energy if they are to remain  bound 
 despite a star formation efficiency as low as 30\% or lower 
suggested by observations. The average local virial ratio $\langle\sigma^2/|\phi|\rangle$  
 is introduced to classify bound clusters as function of their critical  
 $\epsilon$ for dissolution. Clusters  dissolving at lower 
 $\epsilon$ achieve the lowest ratio.  We applied this classification parameter 
successfully to Michie-King and Hernquist-type distribution functions. The Plummer 
sphere is exceptional in that it defies this and other classification parameters we tried. 
 The reasons for the discrepancy include less effective energy redistribution during the 
expansion phase for this case. 
\end{abstract}  

\section{Introduction} 
This article is the second of two addressing the question of 
the survival of bound star clusters which suffer 
mass loss through gas expulsion. Rapid gas removal may unbound a star cluster if the star formation efficiency $\epsilon$ falls below a critical value, where  

 \begin{equation} \epsilon \equiv \frac{M_\star}{M} = \frac{M_\star}{M_\star + M_{gas}}  \label{eq:epsilon} \end{equation}
is the ratio of stellar mass $M_\star$ to total system mass $M$  (cf. Hills 1980; Adams 2000; Clarke et al. 2000). 
 The \sfe\ $\epsilon < 0.4$ at birth   derived observationally on the scales of young clusters (Lada 1999). 
 Using the virial theorem, Hills (1980) found that the gas-free cluster would settle to a new configuration of radius $R_\star$ 
known in terms of the initial radius $R$ from 
\begin{equation} \frac{R_\star}{R} = \frac{1}{2} \frac{M_\star}{M_\star - \half M} = \frac{1}{2}\frac{\epsilon}{\epsilon - \half} \  \label{eq:ratio} \end{equation} 
and hence $R_\star \rightarrow \infty$ if $\epsilon \le \half$, by and large the value of 
reference in this field (e.g., Clarke et al. 2000). 
In Boily \& Kroupa (2002, hereafter Paper I) we have shown that the value $\epsilon$ for which $R_\star \rightarrow \infty$ 
is function of the equilibrium stellar distribution function (\df)  of the cluster at formation. 
 The fraction of stars lost is obtained from the phase-space \df\ $F(\boldv{r},\boldv{v})$
 through the integrals (Paper I)  

\begin{equation} 
1 - \lambda_e = \frac{\displaystyle \int_{v_{e,\star}}^{v_e} F(\boldv{r},\boldv{v}) \,  \rmd^3\boldv{r}\, v^2\,{\rm d}v }{ \displaystyle \int_{0}^{v_e} F(\boldv{r},\boldv{v})\,  \rmd^3\boldv{r}\, v^2\, {\rm d}v} = \frac{\displaystyle \int_{v_{e,\star}}^{v_e} f(v) \, v^2\,{\rm d}v }{ \displaystyle \int_{0}^{v_e} f(v)\, v^2\, {\rm d}v}\, ,
\label{eq:escape} \end{equation}
where $f(\boldv{v}) \rmd^3\,\boldv{v} \equiv F(\boldv{r},\boldv{v}) \rmd^3\boldv{r}\,\rmd^3\boldv{v}$ is the stellar 
velocity \df\ at birth, and $\lambda_e$ the bound fraction of stars {following} gas expulsion. 
 Iterating on (\ref{eq:escape}) to account for escaping stars provides a 
self-consistent selection criterion based on the local escape velocity, $v_{e,\star}$ (see Paper I for further details). 

 The procedure outlined here does not account for the dynamical evolution triggered by the loss of gas, when, on the whole, the 
residual gas mass ($M_{gas}$ above) represents a non-negligible fraction of the total system mass $M$.  In reality 
stars must orbit in space before leaving the system, and they may 
exchange gravitational 
energy with the background potential in the process (cf Binney \& Tremaine 1987 $\equiv$ BT+87 for a discussion), 
when the cluster expands as a result of  the shallower potential well felt by the stars.
 The purpose of this article is to establish the validity of (\ref{eq:escape}) once dynamical 
evolution is fully taken into account. In this respect we continue in the lineage of  Lada, Margulis \& Deardorn (1984) 
 but with the 
benefice of much larger-N simulations than was possible then (tens of thousand compared with hundreds). This allows us to 
 work in the strictly collision-less regime, with much reduced noise in the potential. 
Because we can bridge over to  Paper I, we will 
 focus on a few calculations while exploring a wide range of \df. 

We approach the dynamics through  N-body numerical integration 
as described in the following section. We compare results for Plummer spheres with the 
analytic result of Paper I. We then apply (\ref{eq:escape}) to truncated isothermal (King) \df, as well as 
centrally peaked Hernquist-type models in subsequent sections.  We
 derive limits from the analytic treatment applicable to star clusters, namely 
that embedded 
clusters with an \sfe\ as low as 30\% require a significant fraction of stellar orbits on  low  velocity
 (high binding energy) in 
order to withstand rapid gas evacuation, concurring with the detection of high-mass stars preferentially 
in dense cluster cores (Testi et al. 1997, 1999). We explore alternative mechanisms to form bound open clusters in the 
closing section of the paper. 

\section{Details of the numerical setup} 
 We performed  N-body calculations with the  multi-grid FFT code {\sc superbox} (Fellhauer et al. 2000) 
and 50,000-particle models.  
We have chosen the scale of  N-body models such that the  total mass $M = 1$, $G = 2$, and cutoff radius $ R = 1.25 $ in all cases. The grid resolution and the selection of 
time-step was done such that the equilibria remained stable, in the sense that  concentric 
10\% mass radii remain constant upon integrating for 5 model units of time (from here on we write m.u. for model units), 
corresponding to $ \simeq 6.73 \ \tcr$ 
where the 
crossing time $\tcr$ is defined from the mean density $\bar{\rho} = 3M/4\pi R^3$, 

\begin{equation} \tcr = \left( \frac{3\pi}{4\,G\bar{\rho}} \right)^{1/2}\, \simeq 0.75\ {\rm m.u.}\ . \label{eq:tcr} \end{equation} 
Typical grid resolutions of from $0.005 $ to $0.01$ in the central region proved adequate for the range of models considered.   
  The same remark  applies to all simulations discussed here. 

The nature of the problem 
at hand means we do not have to treat the gas dynamics as such: instead we 
 note that the virial ratio 
\begin{equation} Q \equiv \frac{2T}{-W} = \frac{M \langle v^2\rangle}{GM^2/R} \propto \frac{\langle v^2\rangle}{M/R}\, ,  \label{eq:virial} \end{equation} 
where $T$ and $W$ are the kinetic and gravitational energies, respectively ($Q = 1$ defines virial equilibrium), 
relates velocity dispersion, mass and radius uniquely. Since gas dynamics is 
 only taken into account insofar as it sets 
the stellar velocity dispersion, we use 
(\ref{eq:virial}) to multiply equilibrium particle velocities by the 
factor $\sqrt{M/M_\star}= \sqrt{1 + M_{\rm gas}/M_\star} = \epsilon^{-1/2}$ as defined in (\ref{eq:epsilon}). As a result $Q \ge 1 $ when we replace $M \rightarrow M_\star$ in (\ref{eq:virial}): the super-virial system is dynamically  identical to the problem posed. 

  The excess kinetic energy leads to rapid expansion of the star cluster. 
 We followed the evolution of the mass distribution in two ways. First we 
monitored  concentric spherical shells, each enclosing 
10\% of the total stellar mass. Second, we counted the number of stars that
remained within a volume of linear size twice as large as the initial system 
radius, or $2\times 1.25 = 2.5$  in model units. 
We found both approaches yield similar results. 

In the first instance the 
fraction of bound stars at the end of the calculation is found to 5\% 
accuracy by plotting the Lagrange radii as function of time and determining 
which ones have turned around and settled to finite values. In the second instance, a star 
 whose radial position shifted from $R$ to $2R$ 
 would have suffered a drop in 
 potential from $\phi_\star = GM_\star/R$ to $\phi_\star = GM_\star/2R$ if 
no stars were lost, which corresponds to the mean binding energy per unit 
mass of the initial cluster of radius $R$. 
Thus counting as unbound stars orbiting outside $2 R$ corresponds to removing stars  whose 
 binding energy is not larger than the mean value derived from the gravity of the stars alone. 
Both methods over-estimate the fraction of stars lost since they 
 impose  their own selection of bound orbits by binding energy: consequently, the numerical results 
 provide lower limits to the fraction of bound stars at given \sfe. 
  We allowed stars more time for escape by doubling the integration time of a few calculations. 
In  all cases our results did not vary by significant amounts, so we adopted as standard  
the setup described above. 

\section{Plummer models} 
\subsection{Analytic solution} 
We recall the analytic result of Paper I 
for Plummer spheres before discussing the time-evolution and bound fraction obtained from N-body runs. 
The isotropic \df\ of the $n = 5$ polytropic Plummer sphere  is $F(E) \propto (-E)^{(n+2)/2} = (-E)^{7/2}$. 
Inserting this in (\ref{eq:escape}) leads to the parametric solution 

\begin{equation} 
 \lambda_e (x) =  \frac{2}{\pi} \left( { \sin^{-1} 
x^{1/2} - p(x)\,(x)^{1/2}\sqrt{1-x} }{}\right)\ , \label{eq:lpara} 
\end{equation} 
where $p(x) =  1 -1210/105\, x
+ {2104}/{105}\, x^2 - {1488}/{105}\, x^3 + {384}/{105}\, x^4$  and the parameter $x$  
\begin{equation} 
 x \equiv \lambda_e\, \epsilon\ ; \ x \subseteq [0, 1]\, . \label{eq:defx}
\end{equation} 
The full solution $\lambda_e(x), \epsilon (x)$  follows from (\ref{eq:lpara}), 
known function of $x$, i.e. $ \epsilon(x) = {x}/{\lambda_e(x)}\ .$ 
The solution curve $\epsilon(x) $ reaches a minimum at $ x \approx 0.2252$ when $\epsilon \approx 0.442$. 
Solutions with smaller $\epsilon$ are impossible, which we interpret as indicating total cluster 
dissolution. Note that the solution $\lambda(x)$ is independent of radius, which means that 
the same fraction of stars are escaping at each radius. 

\subsection{N-body calculations} 
 The N-body realisation of a Plummer model is complete once we specify a truncation radius 
and a Plummer length $R_p$. We set $R_p = 0.10$  so that the numerical models contain 99\% of 
the total mass of the model within $r = 1.25$.  Note that the mean density of a Plummer sphere is $(R_p/r)^3 \sim 10^{-3} \times$ the central density so that $\tcr$ corresponds to 
about thirty central crossing times: consequently the timesteps of integration was set to 
$< 1/30$ m.u. to resolve the dynamics at the centre.

The results of runs with $\epsilon$ ranging from 0.4 to 1 
are listed in Table~\ref{tab:plummer} (see also Fig.~2 of Paper I). 
We find the general trend and   quantities in good 
agreement with the results derived from (\ref{eq:lpara}). Specifically, 
 the fraction of bound stars drops very rapidly around $\epsilon = 0.44$; furthermore, we find 
no indication that  stars remain bound for the case where 
$\epsilon = 0.40$. The N-body calculation for $\epsilon = 0.45$ gives a bound fraction 
of $\approx 44\%$. The critical \sfe\ from our numerical models is therefore $42.5 \pm 2.5\,\%$. This is 
in agreement with the findings by others for similar initial conditions  (Lada,
Margulis \& Deardorn 1984; Goodwin 1997; Geyer \& Burkert 2001). We may draw some confidence from the agreement between 
N-body models and the use of different algorithms for integration: for example, 
 Geyer \& Burkert used a TREE algorithm for their SPH-Nbody calculations. \newline 

Table~\ref{tab:plummer} lists the ratios between radii of five concentric spherical 
 shells taken when the system has settled to a new equilibrium, and their respective  
initial radius. Each shell therefore encloses the same Lagrangian mass as it did initially, 
i.e. the same fraction of 
the initial total stellar mass; we therefore refer  loosely to the shells as `Lagrangian radii'. 
The average of these five ratios is given in the 
column noted $R_\star/R_o$ and compared with equation (\ref{eq:ratio}). 
Values derived from (\ref{eq:ratio})  typically over-estimate the 
mean cluster radius, however note that the disagreement with the numerical N-body 
calculations becomes severe only for low \sfe. 
The ratios of Lagrange radii  help assess the morphological evolution of the cluster 
in the following way: If the initial Plummer profile after gas expulsion and dynamical 
relaxation were to settle to an expanded (homologous)  version of itself, the ratios between 
the Lagrange radii at the end of the calculations and their initial values would give the same expansion factor for 
every Lagrange radius. 
At the end of the N-body calculations, in equilibrium, we find this to hold true in the inner 
region of the system and for an \sfe\ $ \gtabout 70\%$.
 We verified  that these fluctuations do not originate from numerical resolution problems: test calculations of 
equilibrium models indicated displacements of mass shells of a few percent at most.    
 At fixed \sfe, the variations between individual ratios do not exceed 10\% up to the 70\% mass shell. 
As the \sfe\ is reduced, the  trend of near-constant expansion factor is gradually  lost, such that 
 when the \sfe\ $\approx 60\%$ or lower the expansion factor increases  monotonically with Lagrangian radii.
 The outermost mass shell expands the most in all the cases, until the \sfe\ becomes so low 
that much of the initial mass leaves the simulation volume. Therefore, the  
new equilibria established by the star clusters differ significantly from the initial Plummer profile, in the sense that its core  expands less than the outer envelope, and is not, as a result, an homologous map of the initial mass profile. 
  (Figure~\ref{fig:Eplummer}[b,c] makes this point graphically, as we discuss below.) 

Equation (\ref{eq:escape})  implicitly assumes that 
the number of stars with energy $E > 0 $ is independent of  details of the time-evolution. Comparisons of bound fractions 
with N-body calculations partly justifies this. 
To see why the iterated selection criterion (\ref{eq:escape}) works, 
we counted the number of stars with positive binding energy $E$ as function of time, as well as the mean radius of two subsets 
of stars: those with negative energy ($E < 0 $), and all the stars within the simulation volume. 

The results are displayed on Fig.~\ref{fig:Eplummer}(a) for a Plummer model with an \sfe\ = 50\% (or, $Q=2$). 
A key feature on this figure is the quick leveling off of the fraction of $E > 0$ stars
at $t \approx 0.2\ m.u.$ (shown as solid line on the figure), while both radii 
expand  until eventually they reach a maximum at $t\approx 0.75$ (dashed lines). 
 The number of $E > 0$ stars increases by a factor $\approx 2.5$ in the course of evolution, indicating a fair 
amount of energy exchanges soon takes place. A similar factor can be computed from (\ref{eq:lpara}) and (\ref{eq:defx}). 
The instantaneous fraction of unbound stars when the gas mass fraction is removed 
 is $1 - \lambda_e(\epsilon = 1/2) = 0.0877$ from (\ref{eq:lpara}), 
while the {\it net} fraction of escapers after iteration  
is $1 - \lambda_e(x) = 0.15$ with $\epsilon(x = 0.425) = 1/2$ 
from (\ref{eq:defx}). The ratio $0.15/0.0877 \approx 1.7 < 2.5$ obtained from the 
 N-body calculation. Nevertheless, the analytic iterative scheme encapsulates the basic dynamics at work during the 
expansion phase.

%

The time-evolution of the system is characterised by the two mean radii (Fig.~\ref{fig:Eplummer}[a], dashed lines). 
At early times, both radii peak simultaneously, however 
they reach different values as might be expected from the selection criterion. This suggests 
 that stars on the whole follow a global radial expansion,
 and hence few $E<0$ stars  have turned around before the expansion phase is over.

The distribution of binding energy provides insight into the speed at which Plummer spheres evolve.  
Fig.~\ref{fig:Eplummer}(b,c) shows two snapshots of the energy distribution at times $t = 1/2$ m.u. and 2, respectively. 
 We have graphed the particles' energy 
 versus their initial energy (after gas expulsion) to highlight evolution.
  The dash triangle isolates a number of stars which had negative binding 
energy at $t = 0$ but now have positive energy due to evolution. 
Note that none of the stars that had positive energy initially acquires
 negative binding energy at later times: indeed their energy at any times parallels the diagonal set by the 
initial conditions, as would be the case for particles cruising away from the cluster centre
 and not participating in the evolution of the central region. 
On Fig.~\ref{fig:Eplummer}(c) a clump forms with high negative binding energy, 
demonstrating  that the cluster is still in its early stages of relaxation at that time. 
Yet the number of stars with $E > 0$ remains practically unchanged at around 11500 from then onwards 
(solid line, Fig.~\ref{fig:Eplummer}[a]). The non-homologous character of the dynamics 
 is confirmed when we consider the changes in 
shape of the distribution of particles at different times.  \newline 

The numerical N-body experiments with Plummer spheres show 
that cluster dissolution takes place at \sfe\ $\approx 42.5\pm 2.5\% $, in good 
quantitative agreement  with (\ref{eq:lpara}). In Paper I, we showed that polytropes of index $n > 5$ 
would dissolve for lower \sfe. However such polytropes are infinite in mass and radius, and hence they 
are of limited use. One exception is the truncated isothermal sphere (Michie-King models, of index $n\rightarrow\infty$) 
which have become benchmarks in globular cluster studies. 

\section{Michie-King models \& an alternative scheme} 

\subsection{Analytic solution} 
 The luminosity  profiles of globular star clusters are well fitted by  
Michie-King models (Michie 1963, King 1966; see Meylan \& Heggie 1997 for a 
recent review). 
 The truncated isothermal \df\ of Michie-King  models takes the form (BT+87, \S 4.4) 

\begin{equation}  F(E) =  \rho_1 (2\pi\sigma^2)^{-3/2}\, \left( \exp\left[\frac{|\phi- \phi(r_t)| - v^2/2}{\sigma^2}\right]  - 1 \right)  \label{eq:dfking} \end{equation} 
for $v \le \sqrt{-2(\phi-\phi[r_t])}$ and $F = 0$ otherwise.  
All Michie-King models are truncated at a finite 
radius, $r_t$, where the density drops to zero. They are differentiated from one another by the 
central ratio 

\[ \frac{\phi[0]-\phi[r_t]}{\sigma^2} \equiv \Psi/\sigma^2\] 
 which is a free parameter.  
 As a function of radius, the mass density $\rho$  flattens out near the centre in all cases.  
The volume of near-constant density bounded by the core radius $r_o$ 

\begin{equation} r_o \equiv \left( 9\sigma^2/4\pi G\rho[\Psi/\sigma^2] \right)^{1/2} \label{eq:defro} \end{equation}
shrinks as $\Psi/\sigma^2$ 
increases. We find on inserting (\ref{eq:dfking}) in the integral (\ref{eq:escape})

\begin{equation} 1 - \lambda_e = 1 - \left. \frac{ 
\exp\left( \displaystyle{\frac{\Psi}{\sigma^2}\hat{\phi}} \right) \left[\displaystyle{ \sqrt{\frac{\pi}{2}}\erf\left(\frac{\hat{v}}{\sqrt{2}}\right) - \hat{v}\, \exp\left(-\frac{\hat{v}^2}{2}\right) }\right] - \displaystyle{\frac{1}{3}} \hat{v}^3 
}
{\displaystyle{ \sqrt{\frac{\pi}{2}} \exp\left( \frac{\Psi}{\sigma^2}\hat{\phi} \right) \erf\left(\frac{\hat{v}}{\sqrt{2}}\right) - \sqrt{2 \frac{\Psi}{\sigma^2}\hat{\phi}}  - \frac{1}{3} \left(2 \frac{\Psi}{\sigma^2}\hat{\phi}\right)^{3/2} } }\, 
\right|_{\, \displaystyle{\hat{v} = \sqrt{2\frac{\Psi}{\sigma^2}\hat{\phi}_\star}}}
^{\,\displaystyle{\sqrt{2\frac{\Psi}{\sigma^2}\hat{\phi}}}} \label{eq:king} \end{equation} 
where $\phi_\star$ is the gravitational potential of the stars alone  and we have used dimensionless variables 

\[ \hat{\phi}(r)  \equiv  \phi(r)/\Psi\, ;\,  \hat{v}(r) \equiv v(r) / \sigma\, . \]
Note that $1 - \lambda_e(\hat{\phi})$  depends on the local potential, and hence it is a function of radius.  
To compute the net fraction of bound stars, we must
therefore re-compute the potential numerically for each evaluation of
$\lambda_e$ in (\ref{eq:king}).  This poses no problem since
$\phi(r\rightarrow\infty) \rightarrow 0$, and the density is known at
each step (though it is no longer a King-Michie profile). 
The integral (\ref{eq:king}) was 
evaluated from standard algorithms (Press et al. 1992, p. 213)
while the potential was computed by mid-point averages 
on a mesh of from 3000 to 6000 points, which 
proved sufficiently accurate for convergence of the results and in particular to 
identify the critical \sfe\ for dissolution. 
Convergence was defined as variations $\delta(1-\lambda_e)$ averaged over the mesh smaller than one 
part in a million for two successive iterations. Results are shown on Fig.~\ref{fig:rho} for an $\Psi/\sigma^2 = 9$ King model, where we graphed the density (top panel) and bound fraction $\lambda_e$ as function of 
 radius for three different 
values of the \sfe, $\epsilon$. The radial dependence of $\lambda_e$ is accentuated for lower values of the \sfe, 
 however it always decreases monotonically with radius until it levels off again near $r = r_t = 1.25$. For an \sfe\ = 45\%, we find $\lambda_e (0) \approx 0.90$ at the centre, and $\lambda_e(r_t) \approx 0.40$ near the edge. The 
profile  $\lambda_e(r)$ remains flat inside $r = r_o \approx 0.013$ for this model, independently of the \sfe. 
 The core region of King models is therefore more robust against dissolution. 
By contrast, the solution for the Plummer model  remains flat at all radii (Fig.~\ref{fig:rho}[a]). 

\subsection{N-body calculations} 
We compared the analytic prediction (\ref{eq:king}) with 
numerical N-body calculations for an $\Psi/\sigma^2 = 9 $ model. 
Our choice of parameter is motivated by 
 the fact that the binding energy of Michie-King models peaks at approximately this value (all models of constant mass 
scaled to the same truncation radius $r_t$), which should, therefore, 
provide the best hope to preserve a bound cluster despite low \sfe. \newline 

 Numerical N-body calculations were done similarly to the case of Plummer spheres, taking care to 
adjust the grid resolution to achieve the same mass resolution per mesh. This way we did not 
bias the calculations against the more centrally peaked Michie-King models. The scales of time, 
mass and length are the same in both the cases. 

Fig.~\ref{fig:king}(a)  graphs the bound mass fraction as function of the \sfe. 
 The dash is the analytic result (\ref{eq:king}), while the open circles show the N-body results.
  For an \sfe\ = 50\%, we integrated one more N-body model up to 
$t = 25\,$ m.u. to allow for a possible larger number of stars to escape, which 
would bring down the bound fraction. The two open circles seen at 
\sfe\ = 50\% on Fig.~\ref{fig:king}(a) differ by a small amount however, 
at $\lambda_e = M^b_\star/M_\star = 0.64\ (t = 5) $ and $ 0.59\ (t = 25)$, respectively; this  gives indirect confirmation of reduced evolution for $t > 5$ and boosts confidence in the analysis. The dashed curve 
 does not give a good fit to the N-body data for the full range of \sfe. We  may understand the reason for this 
discrepancy by comparing the evolution of stars with negative energy to the cluster as a whole, as done for the Plummer 
sphere. Fig.~\ref{fig:Eking} graphs 
the bound-stars mean radius (short-dash), cluster mean radius (long-dash) and fraction of stars with $E > 0$ (solid) 
for the 
case when the \sfe\ = 50\%. Comparison with Fig.~\ref{fig:Eplummer} shows that now the cluster 
  mean radius continues to expand rapidly, while  the fraction of stars 
with positive energy and the $E<0$ stellar mean radius flatten out on the same (short) timescale.  
%
The relatively smooth distribution
of stars with  $E < 0$ in these early stages of evolution is  a strong
 indication that the bound stars have already settled (mixed) in the inner region (Fig.~\ref{fig:Eking}[b,c]). 
Furthermore, the lack of sharp features in the $E>0$ quadrant indicates that the motion 
of the stars involves more orbit-crossing (or, mixing) than for the Plummer model. 
 In other words, the assumption that the fraction of bound stars can be identified from (\ref{eq:king}) 
 independently of the dynamics is now invalidated, and we must seek a remedy. 

\subsection{An improved scheme} 
Our hint comes from Fig.~\ref{fig:Eking}. Since the dynamics of the bound stars settles quickly compared with the bulk, we deduce 
that these stars virialise on a very short time scale. We idealise the situation and make this timescale infinitesimal, as for 
the gas-evacuation timescale of the problem. Then 
a dimensional argument relates the mean square velocity $\leftb v^2\rightb$ to 
the self-gravitating mass $M$ and radius $R$, 

\begin{equation} \leftb v^2 \rightb \propto \frac{GM}{R} \propto \phi = \phi_\star + \phi_{\rm gas} \ .\label{eq:virialmean} \end{equation} 
 Since the escaping stars are removed quickly, and the bound stars evolve on their 
own timescale, as hinted from Fig.~\ref{fig:Eking}, 
we may obtain a better estimate of the bound fraction if we take account of the self-gravitating 
energy of the bound stars alone when defining the velocity \df. Thus let 

\[ \phi_\star \rightarrow \phi_\star(E<0)  \] 
and truncate the \df\ at the escape velocity $v_e$ obtained from 

\begin{equation} \leftb v^2_e \rightb = 2 \times \left( \phi_\star(E<0) + \phi_{\rm gas} \right)\ . \label{eq:newve} \end{equation} 
The transformation is inexact but useful, since it meets one's expectation  that escaping stars in a Michie-King model will come from the outer reach of the cluster (cf. Fig.~\ref{fig:rho}),
 and hence 
the transformation (\ref{eq:newve}) corresponds to the limit of a self-gravitating bound core de-coupled from a loose envelope. 
 We may deduce this also from the correlation in energy seen on Fig.~\ref{fig:Eking}(b,c), where stars with initially high-binding energy remained the most bound  at later times. 
 
We repeat our iterative procedure with (\ref{eq:king}) but with the bounds of integration defined from (\ref{eq:newve}) so that  
 
\begin{equation} 1 - \lambda_e = 1 - \left. \frac{ 
\exp\left(\displaystyle{ \frac{\Psi}{\sigma^2}\hat{\phi}} \right) \left[\displaystyle{ \sqrt{\frac{\pi}{2}}\erf\left(\frac{\hat{v}}{\sqrt{2}}\right) - \hat{v}\, \exp\left(-\frac{\hat{v}^2}{2}\right) }\right] - \displaystyle{\frac{1}{3}} \hat{v}^3 
}
{\displaystyle{ \sqrt{\frac{\pi}{2}} \exp\left( \frac{\Psi}{\sigma^2}\hat{\phi} \right) \erf\left(\frac{\hat{v}}{\sqrt{2}}\right) - \sqrt{2 \frac{\Psi}{\sigma^2}\hat{\phi}}  - \frac{1}{3} \left(2 \frac{\Psi}{\sigma^2}\hat{\phi}\right)^{3/2} } }\, 
\right|_{\, \displaystyle{\hat{v} = \sqrt{2\frac{\Psi}{\sigma^2}\hat{\phi}_\star(E<0)}}}
^{\,\displaystyle{\sqrt{2\frac{\Psi}{\sigma^2}}(\hat{\phi}_\star(E<0)+\hat{\phi}_{\rm gas})^{\frac{1}{2}} }} \label{eq:newking} \end{equation} 
The fraction of escapers $1 - \lambda_e$ is now a stronger function of the computed bound stellar potential $\phi_\star(E<0)$.

The bound fraction computed from (\ref{eq:newking}) is displayed as the solid line on Fig.~\ref{fig:king}(a). 
We find now excellent agreement with the numerical 
 N-body results. 
  The better fit means we may  concentrate on Eq. (\ref{eq:newking}) and explore its range of applicability. Our intuition in writing down (\ref{eq:newking}) came from considering the central core as self-gravitating. To see how well the scheme performs for less centrally concentrated 
distributions, we performed a comparison between the analytic scheme and N-body calculations of an Michie-King $\Psi/\sigma^2 = 3$ model. The 
results are shown on Fig.~\ref{fig:king}(b). We have displayed the results obtained from (\ref{eq:newking}) (solid line) as well as the old scheme (\ref{eq:king}) (dashed line). Once more the N-body data falls much closer to the solid line, though the fit is much poorer. The more extended core of the Michie-King $\Psi/\sigma^2 = 3$ means the region of uniform-density stretches closer 
to the edge of the system: the situation is similar to that of harmonic motion, in which case stars which visit the 
edge escape more easily  (Paper I). 

 
Table~\ref{tab:compare} lists values 
of the critical \sfe\ when cluster dissolution would occur,
 for a sequence of six Michie-King models along with  their characteristic radii. The parameter $\Psi/\sigma^2$  
ranges from 1 to 12. The bound fraction $\lambda_e = M^b_\star/M_\star$ is displayed on Fig.~\ref{fig:king}(c), 
where the free parameter $\Psi/\sigma^2$ assumes values of 3, 6, 9 and 10. 
All curves drop to  
zero bound fraction between an \sfe\  $\epsilon \approx 0.523$ for $\Psi/\sigma^2 = 3$, to $\approx 0.401$ for $\Psi/\sigma^2 = 9$. Models with $\Psi/\sigma^2 > 9 $ reach a higher central density, however the mass within the core region is a smaller 
fraction of the whole; furthermore, the half-mass radius of these models increases with   $\Psi/\sigma^2 $ so that more 
stars come close to the system edge: as a result  $\Psi/\sigma^2 > 9 $ would  dissolve increasingly more easily 
(Table~\ref{tab:compare}). \newline

The trend of \sfe\ when dissolution occurs in relation to the core-halo structure of Michie-King models hints that 
the characteristics of the core determine cluster survival. To discover how the mass profile may affect the outcome, 
we extend our iterative scheme to distributions with no harmonic cores. 

\section{Hernquist profile} \label{sec:hernquist} 
 Hernquist (1990) introduced a well-known centrally peaked  $\rho \propto r^{-1}$ 
density profile. The integrated mass is finite but fills the entire space (see 
the Appendix for details). Experiments with N-body calculations show that a Maxwellian 
 velocity distribution, 
\begin{equation} f(\boldv{v})\rmd\boldv{v}^3 \propto v^2 \exp(-v^2/2\sigma^2) \rmd v\, ,
\label{eq:Maxwellian} \end{equation} 
provides a self-consistent 
velocity field everywhere (Hernquist 1993; Boily et al. 2001). 
Inserting (\ref{eq:Maxwellian})  in (\ref{eq:escape})
yields

\begin{equation} 1 - \lambda_e(\Upsilon_\star) = 1 - 
 \frac{ \sqrt{\Upsilon_\star}\, \exp(-\Upsilon_\star) - \erf(\sqrt{2\Upsilon_\star})
   }{\sqrt{\Upsilon} \exp(-\Upsilon) - \erf(\sqrt{2\Upsilon}) }, 
\label{eq:hernquist}
\end{equation} 
where the dimensionless potential $\Upsilon(r)\equiv \phi/\sigma^2 (r)$ and
$\erf(x)$ is the error function.  
 The relation (\ref{eq:hernquist}) is once more 
  a function of the local stellar potential, $\Upsilon_\star = \phi_\star/\sigma^2$. A  run of 
bound fraction $\lambda_e$ decreases with decreasing $\Upsilon_\star$; $\lambda_e (\Upsilon_\star = 0 ) = 0 $ 
 if we hold $\Upsilon$ constant in (\ref{eq:hernquist}).

To implement the scheme, we truncated the mass profile at a radius $ r = 200\ r_c$, when the spherical 
volume encloses some $\approx 99.5\%$ of the total system mass.
 Errors in binding energy proved completely negligible for our purpose. 
The integral (\ref{eq:hernquist}) was 
evaluated as for the King models. Convergence was once more defined 
as variations in $\delta(1-\lambda_e)$ smaller than one 
part in a million for successive iterations. 

The result of iterating on  (\ref{eq:hernquist}) 
is shown as  the solid curve on Fig.~\ref{fig:hernquist_all}. 
 The curve follows a pattern similar to the solution for Plummer or King models but  breaks at a lower \sfe.  
 For example, the fraction of bound stars is still $>  50\%$  for an \sfe\ = 40\%.   We find  that 
total  cluster disruption  would take place in this case at a critical \sfe\ = 35.3\%, well below 
the prediction (\ref{eq:ratio}) and results for Plummer models. 

The velocity dispersion of an Hernquist model 
drops at the centre of the system as seen when taking the limit $\sigma(x \rightarrow 0) = 0  $ in (\ref{eq:hernquist_sigma}). We would therefore anticipate 
a smaller fraction of stars escaping from the central region where $\rho \propto r^{-1}$ compared with the 
outer envelope where $\rho \propto r^{-4}$. On Fig.~\ref{fig:hernquist}(a), we graph the density profiles for
 Hernquist models and three representative values of the \sfe. The equilibrium profile (\sfe\ = 100\%) 
is shown as the solid curve on the 
figure. The bottom panels graph the fraction of bound stars $\lambda_e = \rho^b_\star/\rho_\star$ as a function of radius. Clearly a greater fraction of stars 
escapes at large radii than at small radii, 
similarly to Michie-King models (cf. Fig.~\ref{fig:rho}).  The fraction of bound stars at 
the centre $\lambda_e(r\rightarrow 0) = 0.983 \approx 1$ for an \sfe\ as low as 37\%, while  $\lambda_e(r\rightarrow \infty) \approx 0.40$ so the ratio $> 2$ between the two limits. 
Both Hernquist and Michie-King models  contrast with the  $\lambda_e = $ constant Plummer solution.  

\subsection{Other, similar profiles: isochrone and Jaffe models}
Contrary to Michie-King and Plummer  models, the Hernquist profile is cuspy in the central region. Is the lower 
\sfe\ for dissolution obtained a result of the  central cusp? In order to understand 
the relative importance of the central cusp, we explored the properties of 
mass profiles that are identical to 
the Hernquist model at large distances, but differ in the central region. 
We have investigated the properties of two 
models, the isochrone and Jaffe models, which, like the Hernquist profile, have a continuous 
density profile (no truncation) but  finite total mass.  Details of each model are given in the Appendix. 
 We note here that the density profile of the isochrone equilibrium remains flat in the centre, 
while the Jaffe model is more steep than the Hernsquist model there, $\rho_J(r\rightarrow 0) \propto r^{-2}$.
 All three show a power-law density $\rho \propto r^{-4}$ at large distances and we chose the scales of length so that each model has the same total mass and density at infinity. 

Results are shown for the isochrone (dotted) and the Jaffe (dashed) profiles together with the Hernquist model 
on Fig.~\ref{fig:hernquist_all}. It is striking that the \sfe\ when dissolution occurs hardly differs between these 
 three cases, despite different central density profiles. 
We compute for the Jaffe model an \sfe\ =  36.9\% and for the isochrone an \sfe\ = 36.4\% for total 
dissolution, which compare well with  the Hernquist case (35.3\%). About 25\% of the total mass falls within the 
central region (inside one length scale) in all three cases. 
  This  suggests that details of the central mass profile matter little when the mass fraction within one core length 
is comparable in each case. The fact that the 
\df\ from which each model derives is a different function of energy (see Appendix) also points to a weak 
dependence of the critical \sfe\ on the shape of the velocity distribution function. This point is 
 reinforced if we recall that the Maxwellian distribution (\ref{eq:Maxwellian}) used is not the one derived 
from Hernquist's (1990) solution. 
 
\subsection{Comparison with  Michie-King \& Plummer models} 
  Our experience with Hernquist-type profiles  suggests that we should concentrate on averaged 
values and not dwell on the detailed profile of the velocity field or mass. 
The virial theorem is an obvious reference quantity. 
 Since self-consistent models fulfill the scalar virial theorem individually, we would compute 
for each an absolute  ratio of total kinetic to gravitational energy = 1/2. Because this 
holds for equilibrium 
 systems as a whole, this ratio will also be obtained from mean values of kinetic  
and gravitational energy. One way to distinguish 
between models might be instead to compute a `local' virial ratio, taking the average of this ratio
 over the entire system. We write 

\begin{equation} 
 \left\langle \frac{\sigma^2}{|\phi|} \right\rangle = \frac{1}{M} \int_0^\infty  \frac{\sigma^2}{|\phi|} 4\pi  \rho(r) r^2\rmd r  \label{eq:meanvirial}
\end{equation} 
where the one-dimensional velocity dispersion $\sigma(r)$ is 
 obtained from integrating the Jeans equation (cf. Eq.~\ref{eq:jeans} below). 
We may compute  (\ref{eq:meanvirial}) for 
each one of the models in the previous section and compare with the dissolution \sfe. 

The results obtained from (\ref{eq:meanvirial}) are listed in Table~\ref{tab:compare}. The quantity 
$\langle\sigma^2/|\phi|\rangle$ maps the dissolution \sfe\ for each of the Hernquist, isochrone and Jaffe 
models. For instance, we computed the lowest value $\langle\sigma^2/|\phi|\rangle = 0.1805$ for the Hernquist model, 
 which also allows the lowest \sfe\ before dissolution. The coincidence between the two trends gives 
us confidence when interpreting $\langle\sigma^2/|\phi|\rangle$ as a predictive tool for the 
critical dissolution \sfe. The same approach applied to Michie-King models also finds a proportionality 
between $\langle\sigma^2/|\phi|\rangle$ and the \sfe\ for dissolution (see Table~\ref{tab:compare}). 
\newline 

   
The dimensionless ratio $r_{\half}/\langle r\rangle$ provides a diagnostic for the survival of bound clusters 
 in relation to 
the \sfe\ (Table~\ref{tab:compare}). 
This ratio applied to the family of Michie-King models proved equally reliable as 
$\langle\sigma^2/|\phi|\rangle$ computed from (\ref{eq:meanvirial}) to identify which 
model is more robust against mass loss (Table~\ref{tab:compare}). 
 If we extended this conclusion to the three continuous cases of Section~5 (Hernquist-type 
models), we would deduce, wrongly, that the Jaffe model is the most robust of those three~: it is 
 in fact the least robust (albeit marginally so), while $r_{\half}/\langle r\rangle = 0.169$ 
is easily the smallest value for these models (Table~\ref{tab:compare}; $r_{999}$ is the radius enclosing 99.9\% of the mass). 
Hence the mass profile alone, and by implication the potential, does not provide 
a criterion for survival applicable to a wide variety of cases.   
The average ratio (\ref{eq:meanvirial}) appears a  better diagnostic, at least as far as 
 we are concerned with  models similar in character to one another. \newline 

 The case of the Plummer sphere cautions against applying (\ref{eq:meanvirial}) to arbitrary systems. 
In this case $\langle\sigma^2/|\phi|\rangle = 1/6$ is 
 lower than that computed for the isochrone, Hernquist or Jaffe models, yet its dissolution \sfe\ is larger than that 
of any of these three models. 
It is worth pointing out that at large \sfe, the Plummer 
sphere actually is {\it more} robust than any of the other models considered (see Fig.~\ref{fig:king}[c,d]). 
 The swift transition to low $M_\star^b/M_\star$ bound fraction and eventual dissolution obtained  for the Plummer may instead 
indicate peculiarities in the dynamics of expansion for this model: we noted that the central volume of Michie-King models 
 re-collapses on itself on a short timescale, while the Plummer solution does not (Figs.~\ref{fig:Eplummer} and \ref{fig:Eking}).

\section{Tidal and collisional effects} General comments are possible  based on data listed 
in Tables~\ref{tab:plummer} to \ref{tab:king2}. These list the factors 
by which concentric shells expanded in the N-body experiments 
 carried out. Clearly the equilibrium clusters that ensued suffered much expansion and mass loss. 
The final equilibrium mass is found from integrating 

\[ M_\star = \int_{V_\star} F(\boldv{r},\boldv{v})\, \rmd^3 \boldv{r} \rmd^3 \boldv{v} \propto \langle f \rangle r_\star^3 (v_\star^2)^{3/2} \] 
where $\langle f\rangle$ is averaged over the entire phase-space volume $V_\star$ and 

\[ v_\star^2 \propto \frac{GM_\star}{r_\star} \] 
since the cluster achieved a new virial equilibrium by hypothesis. Isolating for $\langle f\rangle $ 
we find 

\[ \langle f \rangle \propto M_\star^{-1/2} r_\star^{-3/2} \ . \] 
From e.g. Table~\ref{tab:plummer}, we find for an \sfe\ = 50\% a bound stellar  mass of $\approx 31\%$ of 
the total initial mass, while 
the radius now expanded on the mean by a factor $\sim 5$, relatively large compared with, e.g., Scorpius data, which suggest an expansion factor of a few between clusters of different ages (Brown 2001). 
Inserting this in the relation above, we 
find a net decrease in the mean phase-space density by a factor $ \sqrt{3}/ 5^{3/2} \sim 6^{-1}$.  The 
 rough derivation shows that although a star cluster may survive gas loss, its binding energy will be much lowered 
and therefore the cluster will be 
subject to significant heating from an external tidal potential. None of the \df's considered here account for 
this heating. This is surely at work in the 
case of open clusters in the Galaxy. Care must be taken when making predictions for 
 the short-term evolution of clusters filling their Roche tidal radius however (Terlevich 1987). 
Baumgardt (2001) finds the orbits of $E \approx 0$ tidally  unbound stars persisting for several 
cluster dynamical times, owing to the shared galactic orbital motion by the 
star and the cluster centre of mass. Thus unbound stars seen in projection against 
the sky may still be counted as cluster members and contribute to the potential energy,  
while the energy criteria we have used assumes instant removal of all stars 
with $E > 0$. This would offset  to a certain degree the net effect of tides due to the large expansion 
 factors seen in the numerical computations. 
Overall, tidal fields will strip clusters of more stars than calculated from (\ref{eq:escape}), however the 
quantities may only be worked out once cluster orbital parmeters are specified. \newline 

The dynamics of Michie-King models following gas expulsion proceeds on a short 
timescale in their centre. We found better agreement with numerical experiments when 
 treating the  stars with $E<0$ initially as a self-gravitating group, 
so resetting the cut-off velocity of the \df\ in the problem posed. By implications, 
the central region undergoes much shell crossing and hence collisional effects there  may yet be of 
importance in this problem.   Kroupa, Aarseth \& Hurley
(2001) evolved an embedded 
 Plummer model with a high-precision direct-summation
$N$-body code with {\em delayed}, but then near-to instantaneous, gas-removal. They 
find a fraction $\simeq 30\%$ of stars remain bound  despite
a low \sfe\ of $\approx 33\% (\epsilon =  0.33)$. 
The results are similar to the analytical results 
for the Hernquist model (Fig.~\ref{fig:hernquist_all}, solid curve), while we would have expected 
complete dissolution for strictly collision-less evolution for this \sfe. The delay in removing the gas  
allows the central part to increase its binding energy by two-body relaxation. The central part therefore 
increases its binding energy at the expense of the outer envelope. The effect 
reported there, combined with the result of the present work, help understand the conditions required to form  
bound objects when the \sfe\ $\approx 20\% $ or lower (Lada 1999).  \newline 

\section{General conclusions} 
 We carried out numerical N-body calculations to verify that 
the iterative 
scheme introduced in Paper I does not grossly underestimate the effects of dynamical evolution:
 We found excellent agreement between numerical calculations and analysis for the case of Plummer spheres. 
Rapid virialisation of the central region
of Michie-King models  required a more careful selection of the 
bound stars from the velocity \df\ for agreement (cf. Eq. \ref{eq:newking}). 
 This confirms our conclusion of Paper I, namely that the initial \df\ of embedded clusters  defines 
the stellar population of bound, gas-free clusters. Some  highlights from the present study are: \newline 

1) The mass-weighted mean ratio $\langle\sigma^2/|\phi|\rangle$ equation (\ref{eq:meanvirial}) provides a 
diagnostic to predict critical \sfe\  for total disruption. Thus systems with a lower value computed from  (\ref{eq:meanvirial}) 
will dissolve at a lower \sfe. We have found this to match the analytic results for all \df\ considered with the exception 
of the Plummer \df\ (Table~\ref{tab:compare}). 

2) When a self-gravitating system evolves such that individual star-star encounters are ignored, any member of the 
Michie-King family of star cluster profile will disrupt if less than 40\% of the gas is turned into 
stars initially.

3) A peaked Hernquist-type profile with decaying central velocity dispersion provides the most 
robust configuration against gas loss amongst the \df\ we considered. We found similar values of critical \sfe\ for 
Hernquist, Jaffe and isochrone \df, which all are predicted to dissolve when the \sfe\ is less than $\approx 35\%$.  

The results obtained here gives a clue to understanding the demographics of massive stars in cluster cores, where the more 
massive stars are found preferentially in denser cores (Testi et al. 1997, 1999): indeed only the more cencentrated clusters 
would survive rapid gas expulsion driven by high-mass stars, while others would disperse. The presence of a population of 
expanding stars around such dense cores would be a strong indication that the clusters underwent such a phase of gas removal. 

\section*{acknowledgments} We are grateful to D.C. Heggie and S.J. Aarseth for comments on a draft version 
of this paper, and to an anonymous referee for constructive suggestions.  
CMB was funded under the SFB~439 program in Heidelberg  and wishes to thank his host R. Spurzem at the ARI for unwavering support.

\section*{Appendix: Details of the \df's} 
In this appendix we give details of the numerical integrations for the \df's 
of the three models with mass densities $\rho(r)$ 
satisfying $\rho(r\rightarrow\infty) \propto r^{-4}$ discussed in Section \ref{sec:hernquist}. 

\subsection*{Hernquist \df}
 The density $\rho$ and potential  of this model vary radially according to 

\begin{equation}  \rho(r\ {\rm or}\ x) \equiv \frac{M}{2\pi} \, \frac{r_c}{r} \, 
\frac{1}{(r+r_c)^3} = \frac{M}{2\pi r_c^3}\, \frac{1}{x}\,\frac{1}{(x+1)^3} =
\frac{1}{2\pi G r_c^2}\, \frac{\phi(x)}{x(x+1)^2}\ , \label{eq:hernquist_rho} \end{equation}
with $r_c$ a free length fixing the point of the power-law turnover, and
$x \equiv r/r_c$. 

The velocity field may be recovered in two ways. 
A useful approximation based on the first moment of Jeans equation (cf. BT+87, \S 4.2; Hernquist 1993) is to evaluate 
the one-dimensional isotropic velocity dispersion from  

\begin{equation} \sigma^2(r) =  \frac{1}{\rho(r)} \, \int_r^\infty \rho(x) \nabla\phi(x) \rmd x  \label{eq:jeans} \end{equation} 
which may be computed  from  (\ref{eq:hernquist_rho}); we find 
\begin{equation} \sigma^2(x) = \phi(x) \, x\, (1+x)^4\, \left\{ \ln \frac{1+x}{x} - 
\frac{1/4}{(1+x)^4} - \frac{1/3}{(1+x)^3} - \frac{1/2}{(1+x)^2} -
\frac{1}{1+x}\right\}, \label{eq:hernquist_sigma} \end{equation} 
which  constrains $f(\boldv{v})$ locally
through $\sigma(r \ {\rm or}\ x)$; we thus have the freedom to choose any profile satisfying
$\sigma(r)$. A Maxwellian velocity \df\ is found to give stable equilibria in N-body
realisations of the model when the \df\ is cut off at the 
escape velocity $v_e$, and $\sigma$ re-normalised to (Hernquist 1993) 

\[ \sigma^2(r) = \hat{\sigma}^2(r) \frac{ \displaystyle{\frac{\pi}{4} {\rm erf}(v_e/\sqrt{2}\sigma) - \frac{v_e}{2\sqrt{2}\sigma} \exp(-v_e^2/2\sigma^2)} }{ \displaystyle{ \frac{\pi}{4} {\rm erf}(v_e/\sqrt{2}\sigma) - \frac{v_e}{2\sqrt{2}\sigma} \exp(-v_e^2/2\sigma^2) - \frac{v_e^3}{6\sqrt{2}\sigma^3}\exp(-v_e^2/2\sigma)} } \] 
such that $ \langle v^2 \rangle = 3 \hat{\sigma}^2 $ as for a  Maxwellian profile.  

The velocity field may be recovered directly from the model \df.  
The \df\ which gives rise to (\ref{eq:hernquist_rho}) is 
obtained directly from an Abel transformation (Hernquist 1990)

\[ F(E) = \frac{M}{8\sqrt{2}\pi^3} \, \frac{1}{r_c^3 v_g^3} 
\left\{ 
 \frac{ \displaystyle{ 3 \sin^{-1} q + q \sqrt{1-q^2} ( 1 - 2q^2) ( 8q^4 - 8 q^2 -3 ) } }{ \displaystyle{ \left( 1 - q^2 \right)^{5/2} } } 
\right\} \]
where the definitions

\[ v_g^2 \equiv \frac{GM}{r_c} \,\, ; \,\, q \equiv \sqrt{ -\frac{E r_c}{GM} } 
         \equiv \left( \tilde{\Phi} - \frac{ \tilde{v^2}}{2} \right)^{1/2}\,\, ; \,\, \tilde{\Phi}(r) \equiv \|\Phi(r)\| \]
apply. The bound stellar fraction $\lambda_e$ is then evaluated from 

\newcommand{\PP}{\tilde{\Phi}}
\[ \lambda_e = \frac{1}{\pi\sqrt{2}} \frac{ \displaystyle{ \int_0^{u} F[E] v^2\rmd v}}{\displaystyle{\PP^4/(\PP-1)}} \] 
where we set $ u = \sqrt{2\epsilon\PP} \le \sqrt{2\PP}$. The indefinite integral then reads 
\[ \displaystyle{ \int_0^{u} F[E] v^2\rmd v} = \]
\[        \left. \left\{ 
          -6\,u^3\,{\sqrt{1 - \PP + \frac{u^2}{2}}}\, \sin^{-1} ({\sqrt{\PP - \frac{u^2}{2}}}) +
                                  \left( -2 + 2\,\PP - u^2 \right) \,
                   \left( 
                           { \left. u\,{\sqrt{\PP - \frac{u^2}{2}}}\,
                            \right[ } 
                                  6\,\PP^4 + 11\,\PP^2\,u^2\,\left( 2 + u^2 \right)  -
                                  \PP^3\,\left( 6 + 17\,u^2 \right)  +
\right.\right. \right.\] \[ \left. \left. \left. 
                                  u^2\,\left( -3 + u^2 + 2\,u^4 \right)  -
                                  2\,\PP\,\left( u^2 + 6\,u^4 + u^6 \right)  
                             { \left]  -
                             3\,{\sqrt{2}}\,\PP^4\,\left( -2 + 2\,\PP - u^2 \right) \,
                             \tan^{-1} (\frac{u}{{\sqrt{2\,\PP - u^2}}}) \right. }
                     \right)  
          \right\}\right/ \left\{ \frac{3}{2}\,
     \left( \PP -1 \right) \,\left( 2 - 2\,\PP + u^2 \right)^2 \right\} \]
The solution $\lambda_e(r)$ is obtained by repeated iterations, upon 
a re-evaluation of the potential. The 
 upper bound of integration is then $u = \sqrt{2\PP_\ast(r)}$, i.e. root 
of the dimensionless escape velocity derived from the stellar potential.  

\subsection*{Jaffe \df} 
The model discussed by Jaffe (1983) derives from the potential-density pair 

\[ \rho_J(r) = \frac{M}{4\pi r_j^3}\, \frac{r_j^4}{r^2 (r+r_j)^2 } \ ; \ 
\Phi_J(r) = \frac{GM}{r_j} \ln \left| \frac{r}{r+r_j} \right| \, . \]
In the above expression $r_j$ is a free scale length.
 Near the centre we find $\rho_J(r\rightarrow 0) \propto r^{-2}$ is 
 diverging like a power of $r$. At large distances 
\[ \lim_{r\rightarrow\infty} \rho_J(r) = \frac{M r_j}{4\pi} \, \frac{1}{r^4} \]
 which is identical to the density of an Hernquist model in the same limit if we 
make $r_j = 2 r_c $. Both systems then have the same total integrated mass, $M$. The velocity dispersion $\sigma(r)$ is known 
explicitly everywhere. Let $x = r/r_j$. Then 

\[ \sigma_J^2(x) = \frac{GM}{r_j} \left\{ 1 - 4x -18x^2 -12x^3  \right\} - 12 x^2 (1+x^2)^2 \Phi(x) \
\raisebox{1ex}{$\underrightarrow{\ x\rightarrow 0\ }$ }\  {\rm constant}. \]
Thus the velocity dispersion levels off near the centre. The \df\ may be expressed explicitly for this 
model (Jaffe 1983; BT+87) 

\[ F_J(E) = \frac{M}{\pi^3 \left( GMr_j\right)^{3/2} } \times \left\{ 
 F_{-}(\sqrt{2 q} ) - \sqrt{2} F_{-}(\sqrt{q}) - \sqrt{2} F_{+} (\sqrt{q}) 
 + F_{+} ( \sqrt{2q} ) \right\} \] 
with $q$ defined as before but with the substitution 
$r_c \rightarrow r_j$. The dimensionless  integral $F_\pm(u)$ is a modified 
Dawson integral 

\[ F_\pm (u) \equiv e^{\mp u^2} \, \int_0^{u} e^{\pm y^2} \rmd y \, . \]
The substitution $F_J(E)$ above in the integral (\ref{eq:escape}) yields no 
 simple expression and we must proceed numerically. The integrals $F_\pm$ however 
  are evaluated efficiently with numerical integrals (cf. Press et al. 1992, \S 6.10). 

\subsection*{Isochrone \df} 
 
The potential-density pair of the isochrone model are (BT+87) 

\[ \rho_I(r) = \frac{ 3a^2(r) (b+a(r) ) - r^2 ( b+a(r) ) }{4\pi a^3(r) (b+a(r) )^3 } \ ; \ 
\Phi_I(r) = - \frac{GM}{b + a(r)} \]
with the definition $a(r) \equiv \sqrt{ b^2 + r^2} $. The asymptotic limit $\rho_I(r\rightarrow 0) \propto $ constant, whereas at large distances 

\[ \lim_{r\rightarrow \infty} \rho_I(r) = \frac{M}{2\pi} \frac{b}{r^4} \ . \]
We therefore recover the same density at large distances for all models if their total integrated 
mass $M$ is equal and the lengths $b = r_c = r_j/2$. In terms of the parameter $q$ the \df\ is known 
explicitly from (H\'enon 1960) 

\[ F_I(E) = \frac{M}{\sqrt{2}(2\pi)^3} \frac{1}{(G M b)^{3/2}}\, \frac{\sqrt{q}}{(2-2q)^4}
\left\{ 27 - 66 q + 320 q^2 - 240 q^3 + 64 q^4  + 3( 16q^2 + 28 q - 9 ) \frac{ \sin^{-1}\sqrt{q} }{\sqrt{q(1-q)}} \right\} \ . \] 
Unfortunately we found no algebraic expression upon inserting $F_I$ in (\ref{eq:escape}) and integrating. Numerical integration presented no difficulty. 

\flushbottom  
\newpage 

\begin{table*}
\caption{Computed quantities for Plummer models together with results of N-body calculations.}
\label{tab:plummer}
\begin{center} 
\begin{tabular}{ccccrccccccc}
 \sfe & $Q$ & \multicolumn{2}{c}{Bound fraction} & \multicolumn{2}{c}{$R_\star/R_o$} & & \multicolumn{5}{c}{Ratio of Lagrangian radii}  \\\hline  
 $100\times\epsilon$     &       &  Eq (\ref{eq:lpara})   &  Numerical   & Eq (\ref{eq:ratio}) &Numerical & & 10\% & 20\% & 30\% & 50\% & 70\% \\ 
      &       &                          &   $\pm 0.05$  &                     & (mean)   \\ 
  100\%  &  1.00  & 1.00  & 1.00 & 1.00 & 1.00 & & 1.00 & 1.00 & 1.00 & 1.00 & 1.00 \\ 
  90\%   & 1.11 &      0.99  & 1.00 &  1.13 & \\
  80\%   & 1.25 &       0.99   & 0.98 &1.33 & 1.37  &  & 1.47 &  1.38 & 1.33 & 1.30 & 1.35 \\
  70\%   & 1.43 &      0.99    & 0.95  & 1.75& 1.68  &  & 1.74 & 1.64 & 1.60 & 1.62 & 1.80 \\
  60\%  & 1.67&        0.95 & 0.86 & 3.00 & 2.40  &  &    2.21 & 2.11 & 2.10 & 2.44 & 3.12 \\
  55\%  & 1.82 & 0.92 & 0.78 & 5.50& 3.39 & &             2.69 & 2.61 & 2.71 & 3.68 & 5.26 \\ 
  50\%   &  2.00 & 0.85 & 0.66  & $\infty$  &   4.35 & & 3.29 & 3.38 & 4.13 & 6.63 & n.a. \\
  45\%   & 2.22 &  0.69       & 0.44 &  &      11.5    & & 5.50 & 10.6 & 18.4 & n.a. & n.a. \\
  40\%   & 2.50 &  0.00       & 0.00  &  & $\infty$ & \\
%
\end{tabular}
\end{center} 
\end{table*}

\begin{table*}
\caption{Computed quantities for a King  $\Psi/\sigma^2 = 9 $  model together with results of N-body calculations. n.a. = non-available.} \label{tab:king}
\begin{center} 
\begin{tabular}{ccccrccccccc}
 \sfe & $Q$ & \multicolumn{2}{c}{Bound fraction} & \multicolumn{2}{c}{$R_\star/R_o$} & & \multicolumn{5}{c}{Ratio of Lagrangian radii}  \\\hline  
  $100\times\epsilon$  &   &  Eq (\ref{eq:newking})   &  Numerical   & Eq (\ref{eq:ratio}) &Numerical & & 10\% & 20\% & 30\% & 50\% & 70\% \\ 
      &       &                          &   $\pm 0.05$  &                     & (mean)   \\ 
  100\%  & 1.00   & 1.00 & 1.00 & 1.00      & 1.00 & & 1.00 & 1.00 & 1.00 & 1.00 & 1.00 \\ 
  90\%   & 1.11   & 0.99 & 1.00 & 1.13      &      &  \\
  80\%   & 1.25   & 0.97 & 0.98 & 1.33      & 1.35  &  & 1.45 &  1.36  & 1.31 & 1.29 & 1.33 \\
  70\%   & 1.43   & 0.91 & 0.92 & 1.75      & 1.65  &  & 1.66 &  1.64  & 1.58 & 1.63 & 1.75 \\
  60\%   & 1.67   & 0.79 & 0.80 & 3.00      & 2.31  &  & 1.96 &  2.10  & 2.15 & 2.34 & 3.00 \\
  50\%   & 2.00   & 0.63 & 0.64 & $\infty$  & 5.56  &  & 3.12 &  3.74  & 4.89 & 10.5 & n.a. \\
  40\%   & 2.50   & 0.27 & 0.39 &           & 7.18  &  & 4.65 &  6.79  & 10.1 & n.a. & n.a. \\
  30\%   & 3.33   & 0.00 &(0.10)&           & $\infty$ &  \\
 20\%    & 5.00   & 0.00 & 0.00 & &  \\
\end{tabular}
\end{center} 
\end{table*}

\begin{table*}
\caption{Computed quantities for a King  $\Psi/\sigma^2 = 3$  model together with results of N-body calculations. n.a. = non-available.}\label{tab:king2}
\begin{center} 
\begin{tabular}{ccccrccccccc}
 \sfe & $Q$ & \multicolumn{2}{c}{Bound fraction} & \multicolumn{2}{c}{$R_\star/R_o$} & & \multicolumn{5}{c}{Ratio of Lagrangian radii}  \\\hline  
 $100\times\epsilon$  &       &  Eq (\ref{eq:newking})   &  Numerical   & Eq (\ref{eq:ratio}) &Numerical & & 10\% & 20\% & 30\% & 50\% & 70\% \\ 
      &       &                          &   $\pm 0.05$  &                     & (mean)   \\ 
  100\%  & 1.00   & 1.00 & 1.00 & 1.00      & 1.00 & & 1.00 & 1.00 & 1.00 & 1.00 & 1.00 \\ 
  90\%   & 1.11   & 0.99 & 1.00 & 1.13      &      &  \\
  80\%   & 1.25   & 0.94 & 0.99 & 1.33      & 1.25  &  & 1.26 &  1.26  & 1.25 & 1.25 & 1.25 \\
  70\%   & 1.43   & 0.86 & 0.97 & 1.75      & 1.55  &  & 1.53 &  1.52  & 1.52 & 1.54 & 1.63 \\
  60\%   & 1.67   & 0.68 & 0.84 & 3.00      & 2.18  &  & 1.93 &  1.95  & 1.98 & 2.13 & 2.93 \\
  50\%   & 2.00   & 0.00 & 0.53 & $\infty$  & 3.09  &  & 2.92 &  3.05  & 3.30 & n.a. & n.a. \\
  40\%   & 2.50   & 0.00 &(0.11)&           & $> 19.00$  &  & $> 19.00$ & n.a. & n.a. & n.a. & n.a. \\
  30\%   & 3.33   & 0.00 & 0.00 &           & $\infty$ &  \\
\end{tabular}
\end{center} 
\end{table*}

\newpage 

\begin{table*}
\caption{ 
Comparison of model characteristics. The \sfe\ at which dissolution occurs is given for 
each model along with mean and half-mass lengths $r_{\half}$ for each case. 
The radius $r_{999}$  within which the mean radius was evaluated defines 
the sphere enclosing 99.9\% of the mass. The characteristic lengths $l$ 
are given for each case. For Michie-King 
models the core length $r_o $ defined by (\ref{eq:defro}) is set such 
that the cluster radius $r_t = 5/4$. 
} \label{tab:compare}
\begin{center} 
\begin{tabular}{cccccccc} 
 King & length $l$  & $\leftb r\rightb$ & $r_t$ & $r_{\half}$ & $r_{\half}/\leftb r\rightb  $  & $\langle \sigma^2/\phi\rangle$ & \sfe  \\ 
  $\Psi/\sigma^2 =$     &  $r_o =$  &  & & & & Eq.~\ref{eq:meanvirial} & $\epsilon$   \\ \hline  
  1   & $0.58$ & 0.463 & 1.250 & 0.440 & 0.950  & 0.808 & 0.529 \\
  3   & $0.20$ & 0.377 & 1.250 & 0.343 & 0.911  & 0.786 & 0.523 \\
 $6$  & $0.10$  & 0.261 & 1.250 & 0.203 & 0.777 & 0.709 & 0.487 \\
 $9$  & $2.50\times 10^{-3}$ & 0.316 & 1.250 & 0.229 & 0.725  & 0.685 & 0.401 \\
 $10$ & $1.25\times 10^{-3}$  & 0.379 & 1.250 & 0.303 & 0.801 & 0.722 & 0.449 \\
 $12$ & $4.98\times 10^{-4}$  & 0.479 & 1.250 & 0.426 & 0.893 & 0.764 & 0.496 \\
\\
 Model & length $l$ & $\leftb r\rightb$ & $r_{999}$ & $r_{\half}$ & $r_{\half}/\leftb r\rightb$ & $\langle \sigma^2/\phi\rangle$ & \sfe  \\  
    &  &  & & & & Eq.~\ref{eq:meanvirial} & $\epsilon$  \\ \hline  
Isochrone & $b = 1$    & 12.83 & 1999.3 & 3.060 & 0.238 &  0.245 &  0.364  \\
Hernquist & $r_c = 1$  & 12.22 & 1998.5 & 2.414 & 0.198 &  0.181 & 0.353 \\
Jaffe     & $r_J = 2$  & 11.83 & 1998.0 & 2.000 & 0.169 &  0.252 & 0.369 \\
\\
Plummer & $R_p = 1$    & 1.925  & 38.714 & 1.305 & 0.678&  0.1667 & 0.442  \\  
\end{tabular}
\end{center} 
\end{table*}

\flushbottom 
\newpage


\setlength{\unitlength}{1in} 
\begin{figure*} 
\begin{picture}(4.5,4.5)(0,0) 
        \put(-.5,0.1){\epsfxsize=0.4\textwidth\epsfysize=0.4\textheight 
\epsfbox{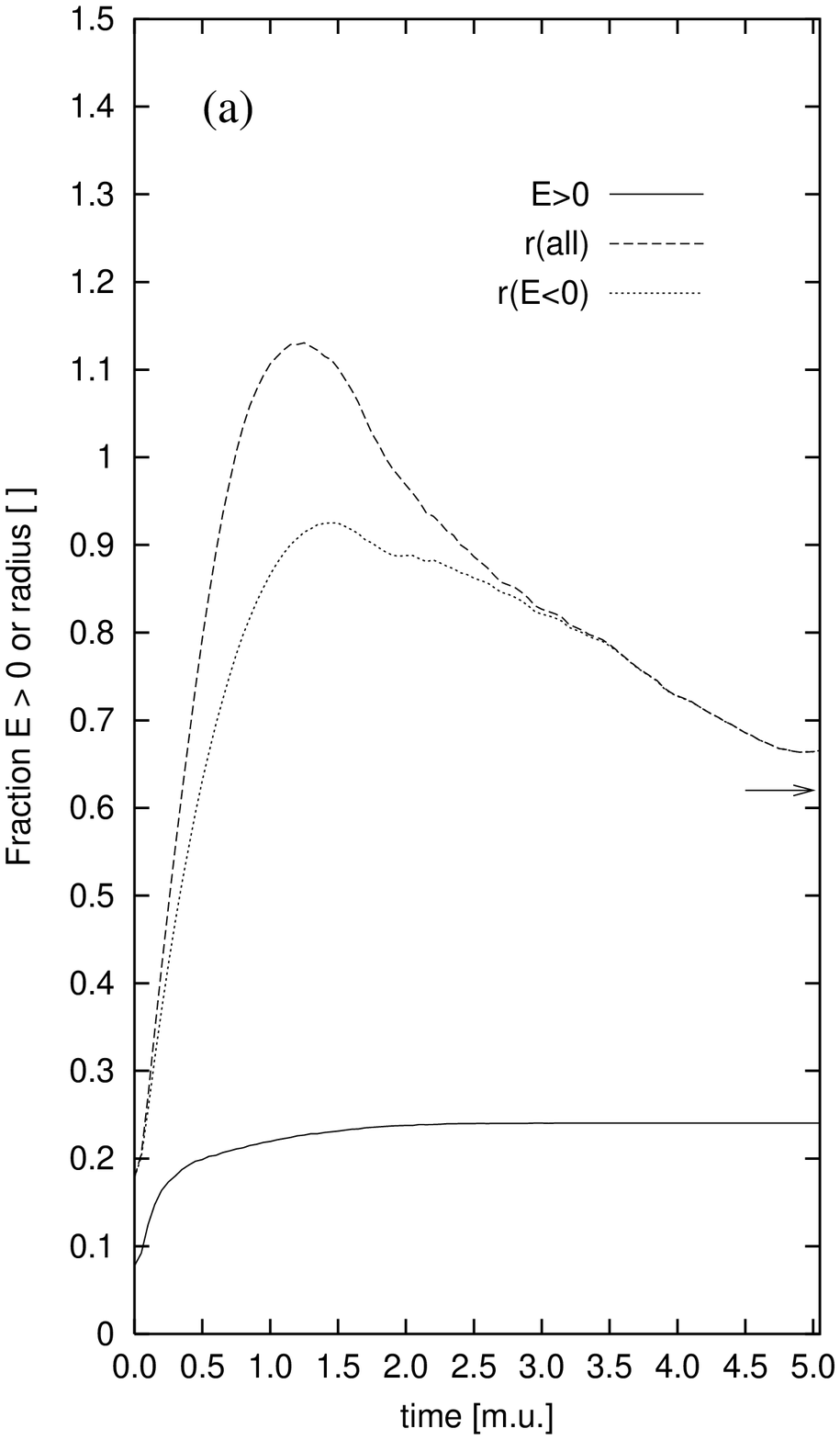}
                  } 
        \put(2.,.1){\epsfxsize=0.43\textwidth\epsfysize=0.43\textheight 
\epsfbox{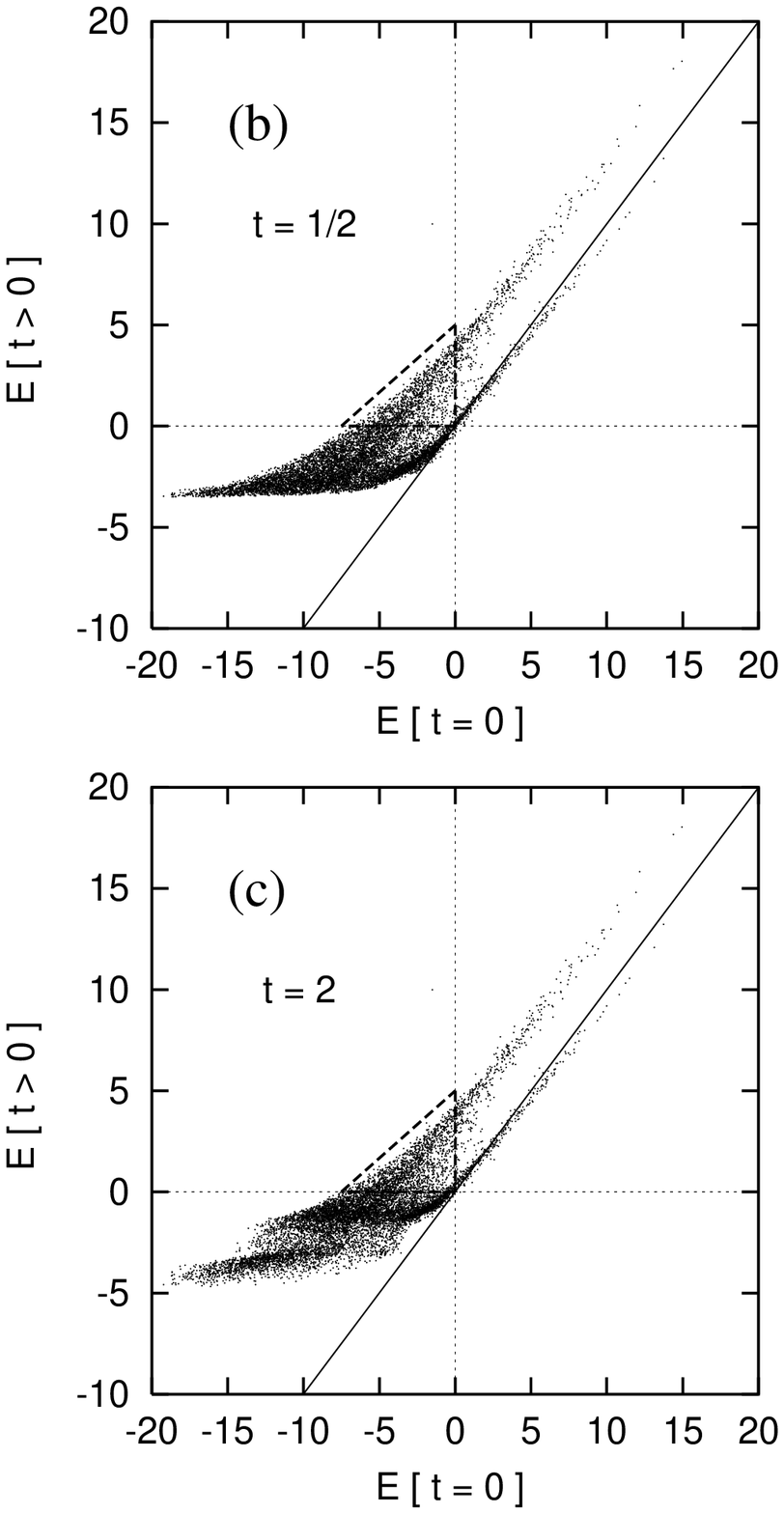}
                  } 
\end{picture}  
\caption{Time-evolution of an N-body calculation. The model Plummer cluster had 
parameters $Q = 2$ and a truncation radius $R = 22\ R_p = 1.25\,$ m.u.; $N = 50,000$ particles were used in the calculation. (a)  
This graphs the mean radius of all stars (long-dash) and stars with $E < 0 $ (short dash) as function of time. (One m.u. of time $\approx 1.3\ \tcr$ defined in [\ref{eq:tcr}].)  
The solid line shows the fraction of stars $N (E>0)/N$  with positive energy. (b) and (c): particle energy versus 
their initial energy (after gas expulsion) at times $t = 1/2$ and $2$, respectively. The triangle indicates stars that had negative 
energy initially which are now unbound.} 
\label{fig:Eplummer} 
\end{figure*} 

\newpage 

\setlength{\unitlength}{1in} 
\begin{figure*} 
\begin{picture}(5,5.)(0,0) 
        \put(-.75,0.1){\epsfxsize=0.5\textwidth\epsfysize=0.5\textheight
\epsfbox{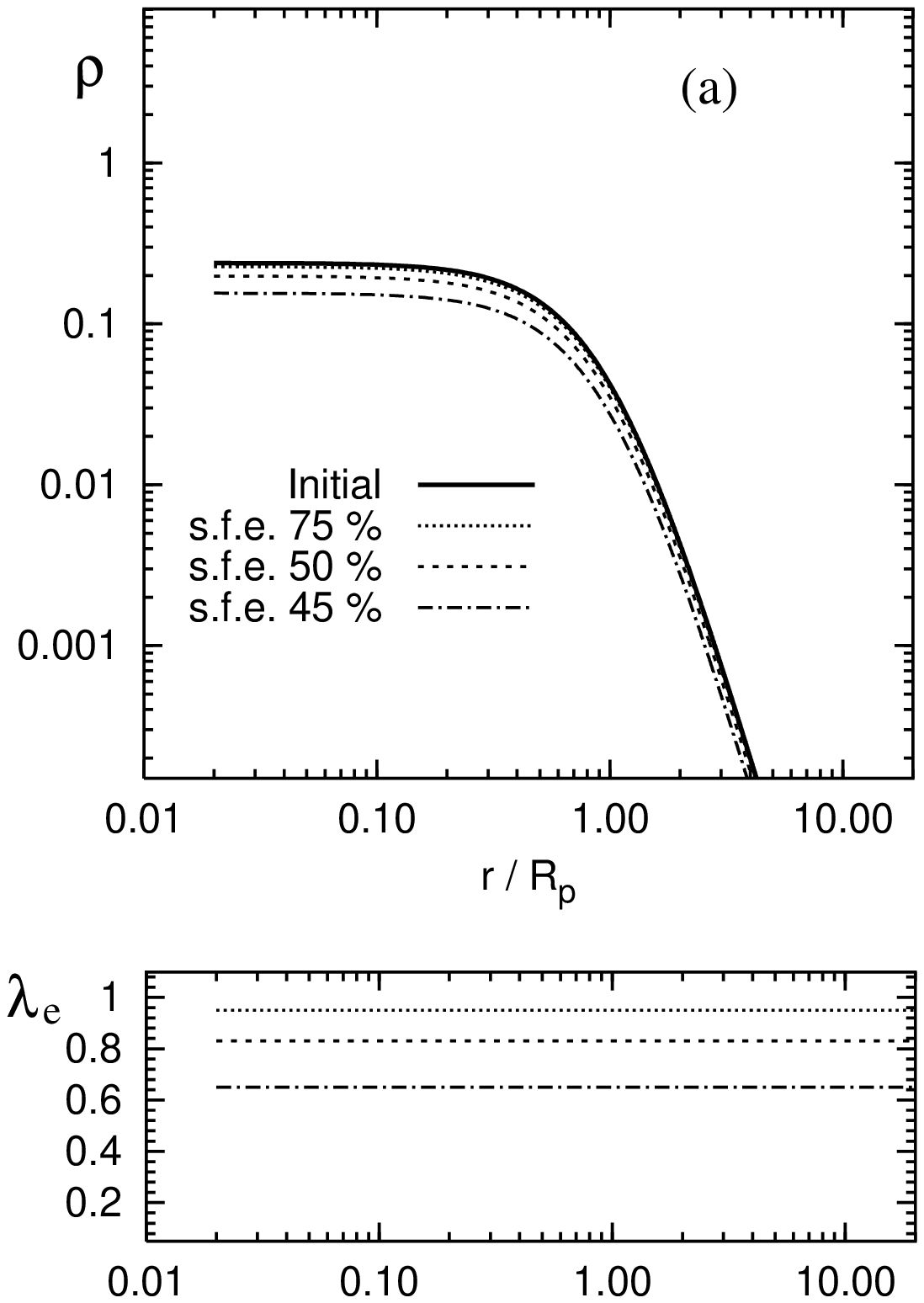}
                  }
        \put(2.5,0.1){\epsfxsize=0.5\textwidth\epsfysize=0.5\textheight
\epsfbox{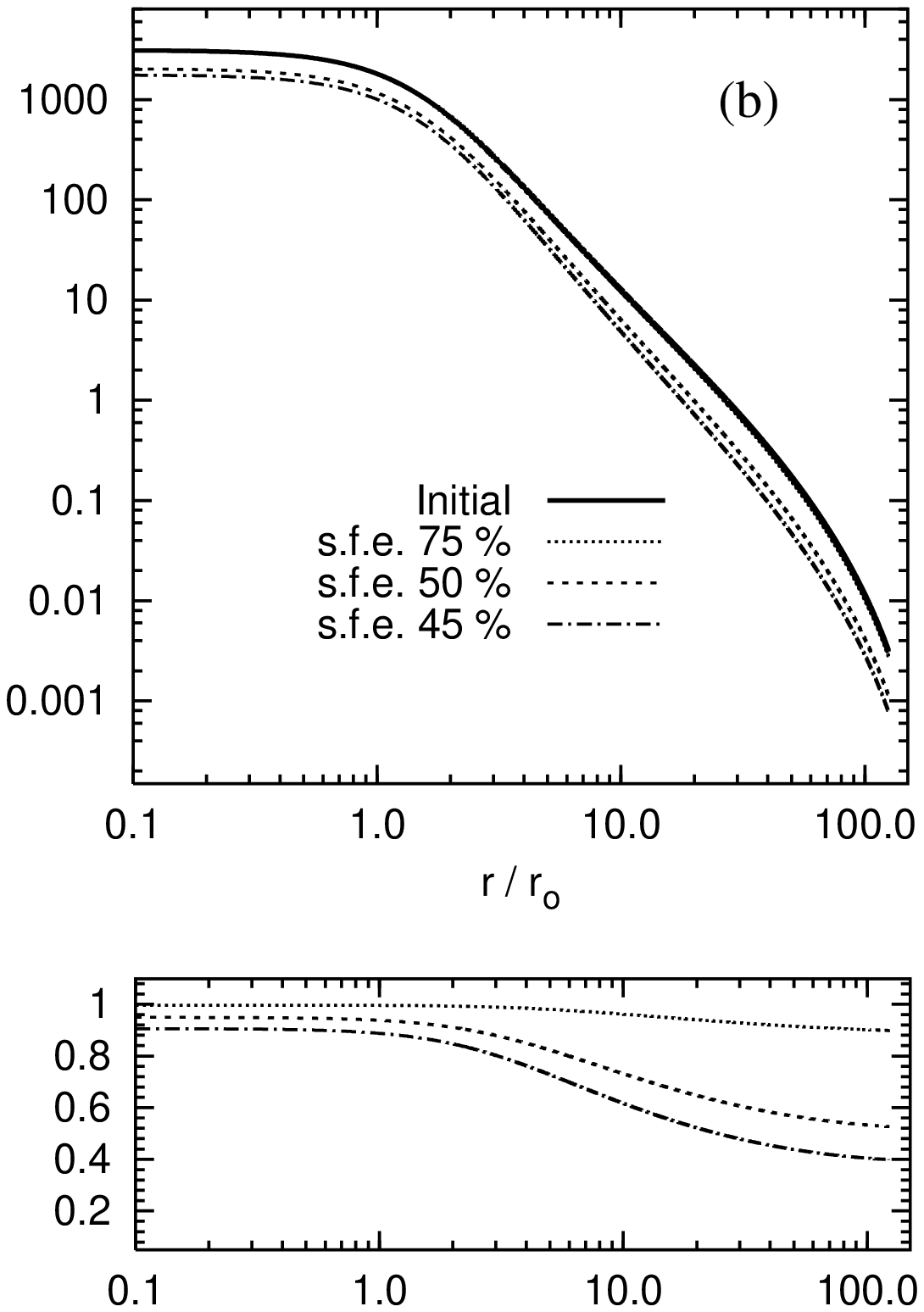} 
                  }
\end{picture}
\caption{Initial and final (after iterations) density profiles for (a) Plummer and (b)
King $\Psi/\sigma^2 = 9 $  models for three values of the \sfe, $\epsilon$. The run of
$\lambda_e$, i.e. the ratio of density to the initial density, is also shown
as a function of radius (bottom panels). Note how the King profile becomes
steeper for small $\epsilon$, while a Plummer model loses mass equally at all 
radii.  } 
\label{fig:rho}  
\end{figure*} 

\newpage 

\setlength{\unitlength}{1in} 
\begin{figure*} 
\begin{picture}(4.5,6.5)(0,0) 
        \put(-1.25,3.2){\epsfxsize=0.45\textwidth\epsfysize=0.45\textheight 
\epsfbox{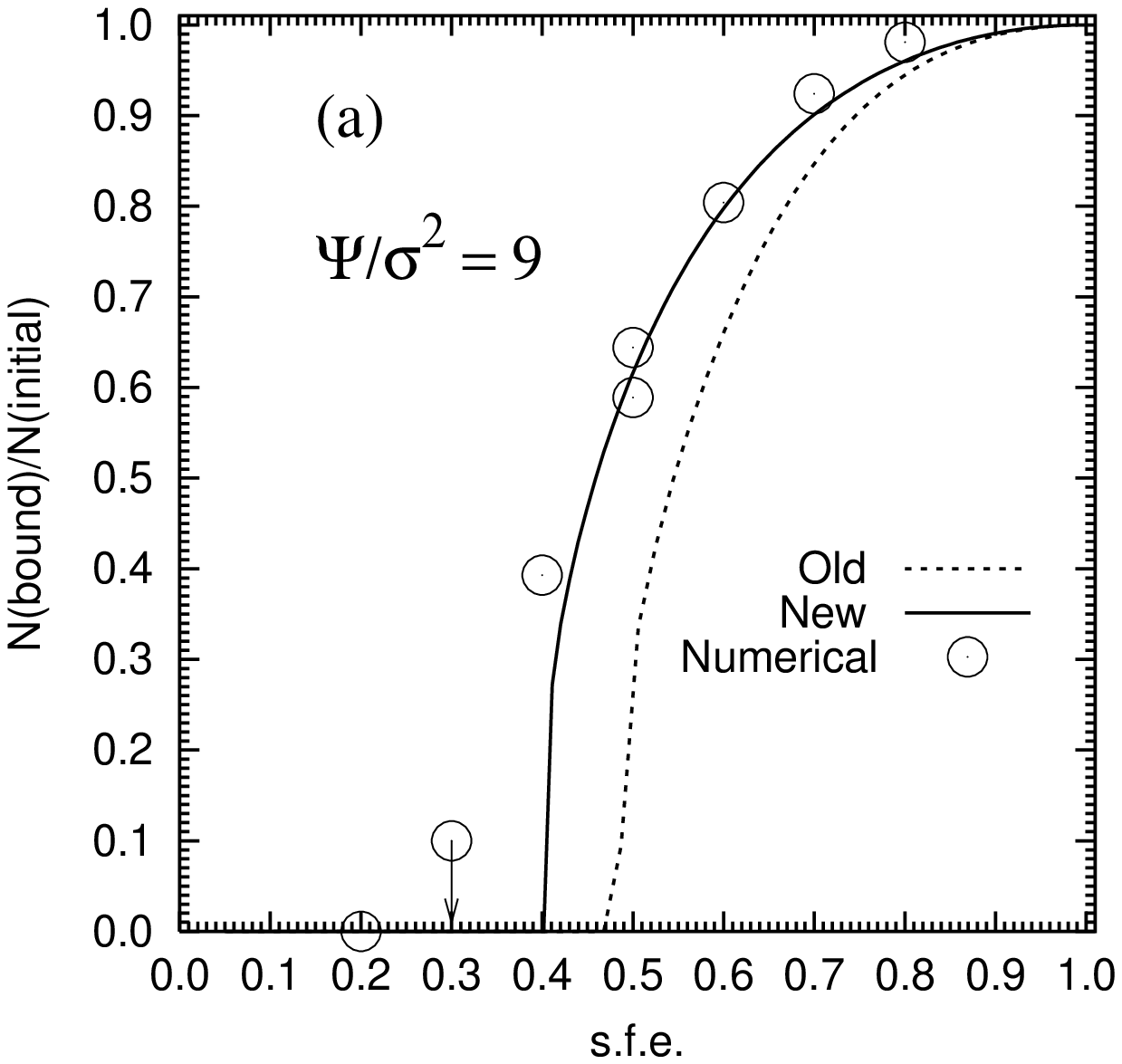}
                  } 
        \put(2.45,3.2){\epsfxsize=0.45\textwidth\epsfysize=0.45\textheight 
\epsfbox{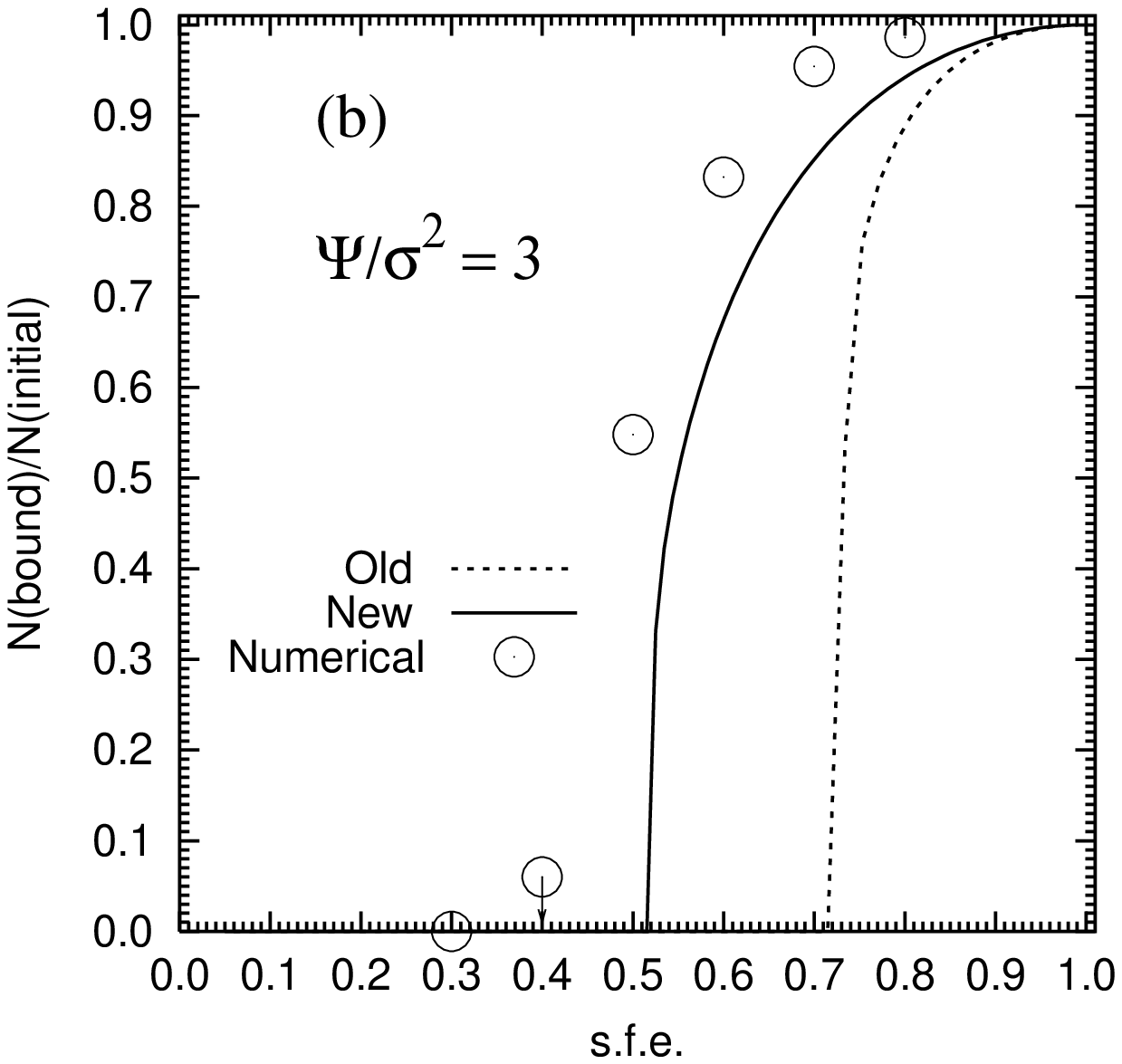}
	}
        \put(-1.25,0.1){\epsfxsize=0.45\textwidth\epsfysize=0.45\textheight 
\epsfbox{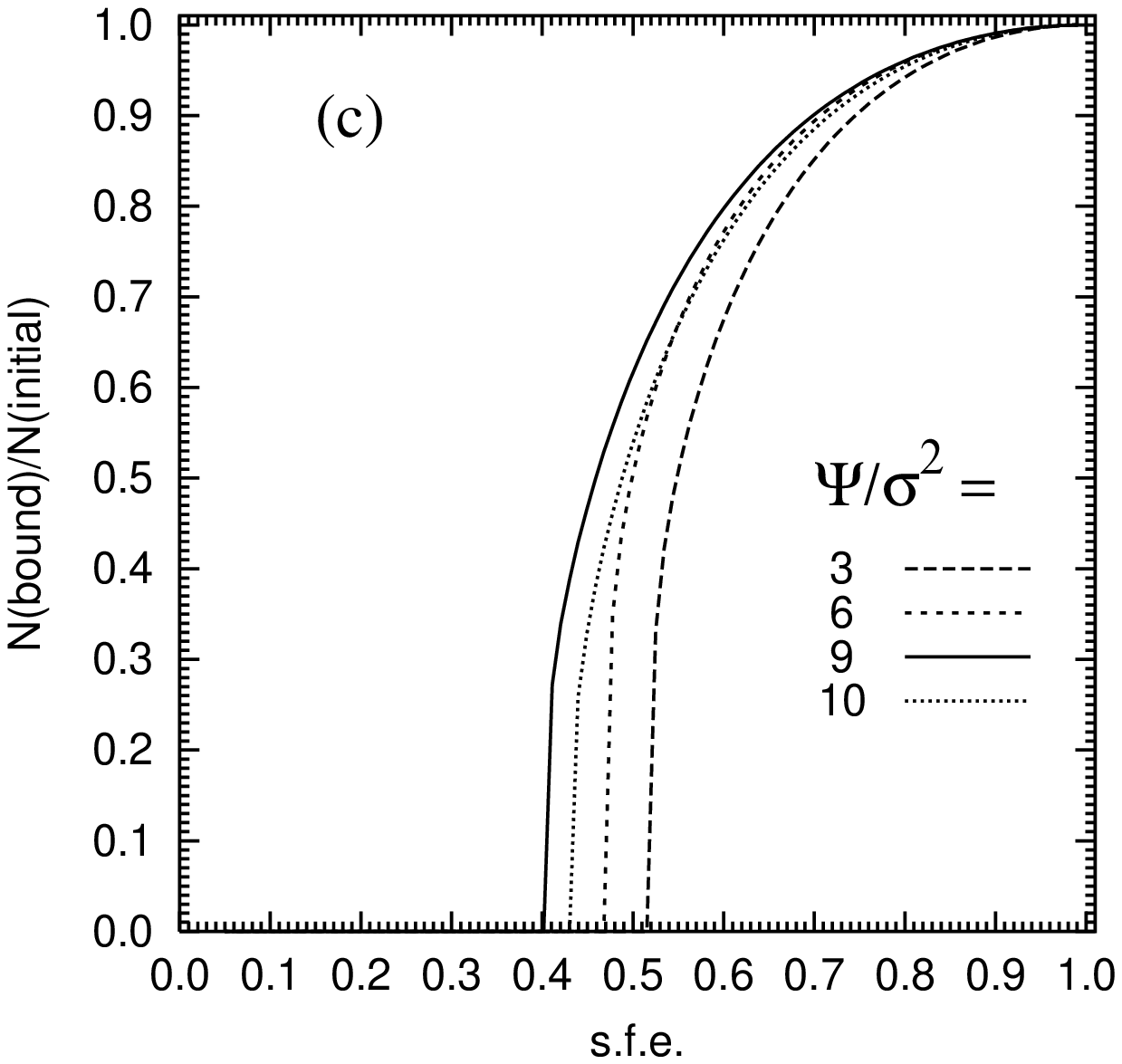}
                  } 
        \put(2.45,0.1){\epsfxsize=0.45\textwidth\epsfysize=0.45\textheight 
\epsfbox{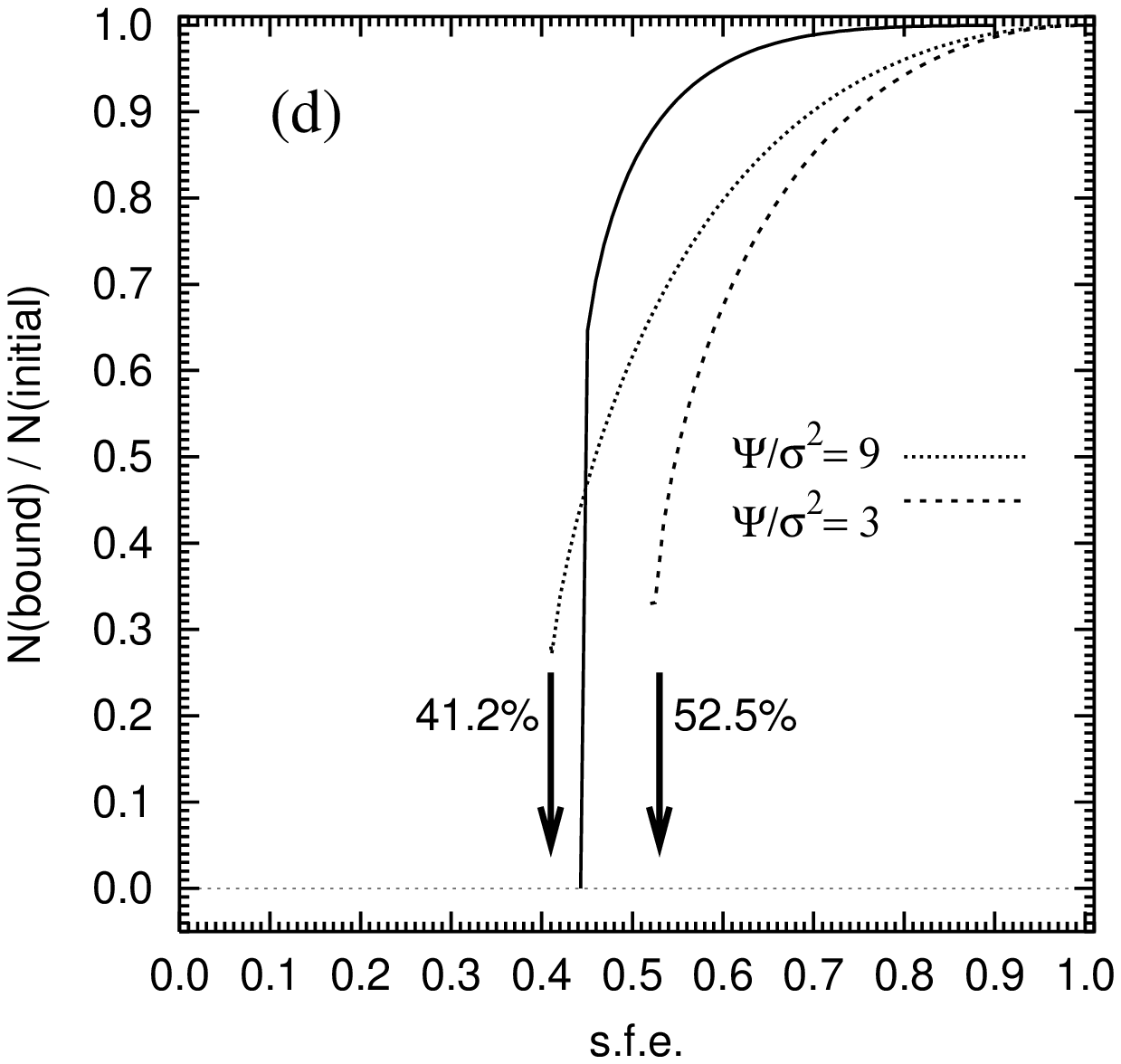}
                  } 
\end{picture}  
\caption{Ratio $ \lambda_e = M^b_\star/M_\star$
 of  bound  
 to initial stars as function of  star formation efficiency $\epsilon$. (a) 
The results for King $\Psi/\sigma^2 = 9$  models derived from (\ref{eq:king}) are shown 
 as the dashed line; the improved solution (\ref{eq:newking}) is the solid line.  
The results of numerical N-body computations are shown as open circles and listed in Tables~\ref{tab:king} and \ref{tab:king2}. 
(b) As (a), but for $\Psi/\sigma^2 = 3$ King models. (c) 
 Solutions for four different King models with $W = \Psi/\sigma^2 = 3, 6, 9 $ and 10. (d) Comparison between two King models and a Plummer model shown as 
solid. The Plummer model dissolves for an \sfe\ below $\approx 0.44$. } 
\label{fig:king} 
\end{figure*} 

\newpage 

\setlength{\unitlength}{1in} 
\begin{figure*} 
\begin{picture}(4.5,4.5)(0,0) 
        \put(-.75,0.1){\epsfxsize=0.4\textwidth\epsfysize=0.4\textheight 
\epsfbox{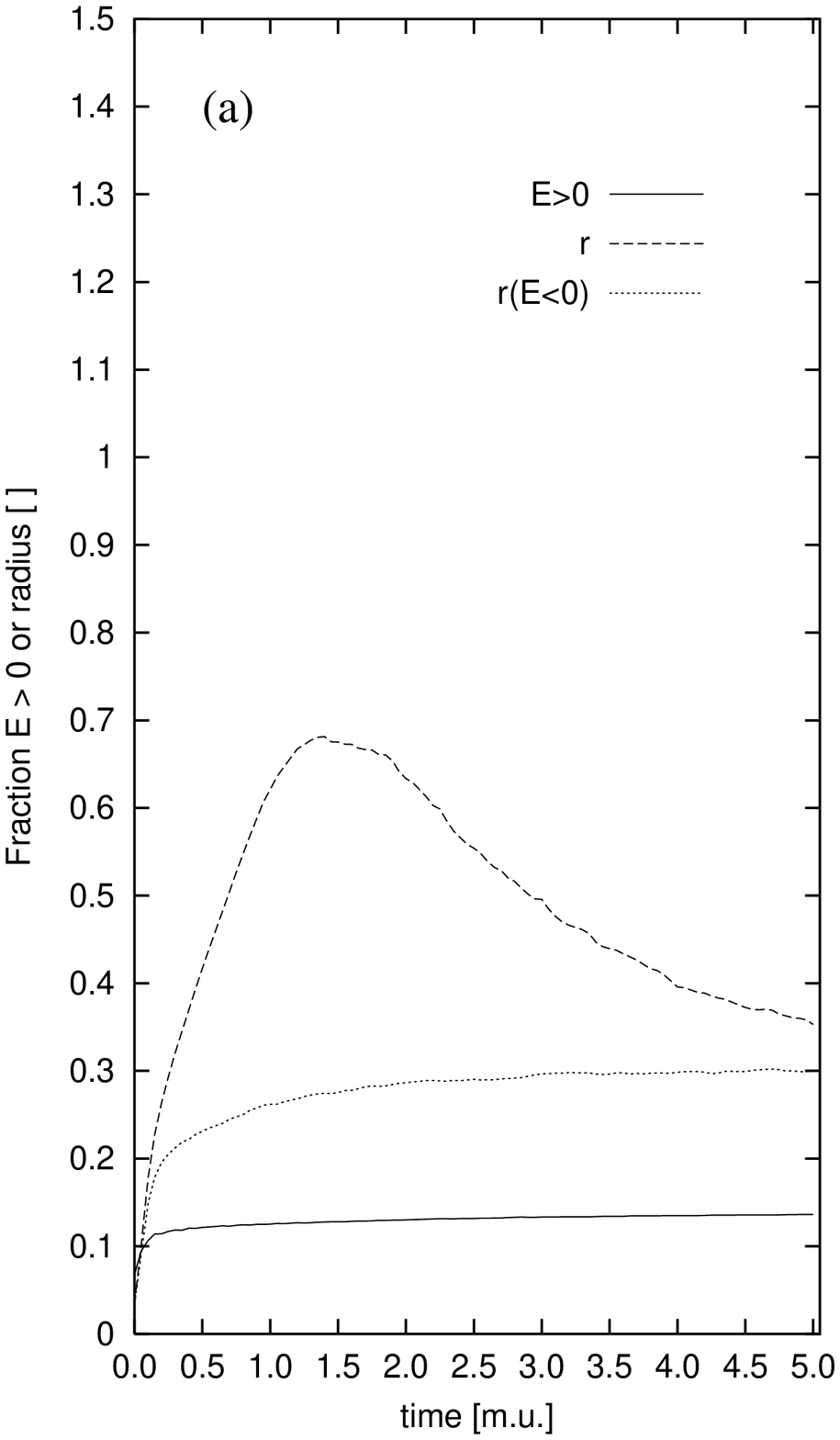}
                  } 
        \put(2.,.1){\epsfxsize=0.43\textwidth\epsfysize=0.43\textheight 
\epsfbox{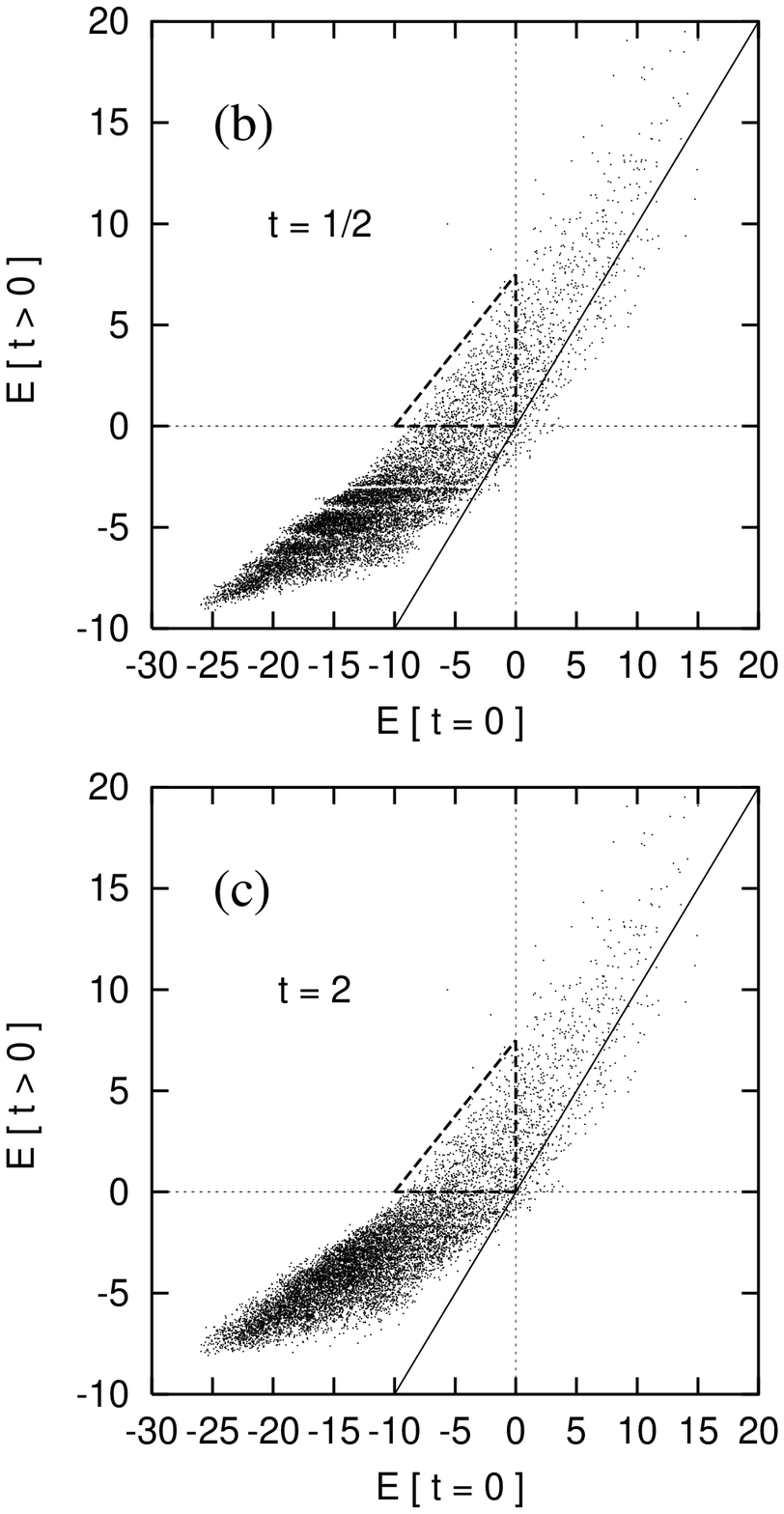}
                  } 
\end{picture}  
\caption{Time-evolution of an N-body calculation. As for Fig.~1 but 
for a Michie-King model with  
parameters $\Psi/\sigma^2 = 9$ and $Q = 2$. (One model unit of time $\approx 1.3\ \tcr$ defined in [\ref{eq:tcr}].) } 
\label{fig:Eking} 
\end{figure*} 

\newpage 

\setlength{\unitlength}{1in} 
\begin{figure*} 
\begin{picture}(5,5.)(0,0) 
        \put(.25,0.1){\epsfxsize=0.75\textwidth\epsfysize=0.75\textheight
\epsfbox{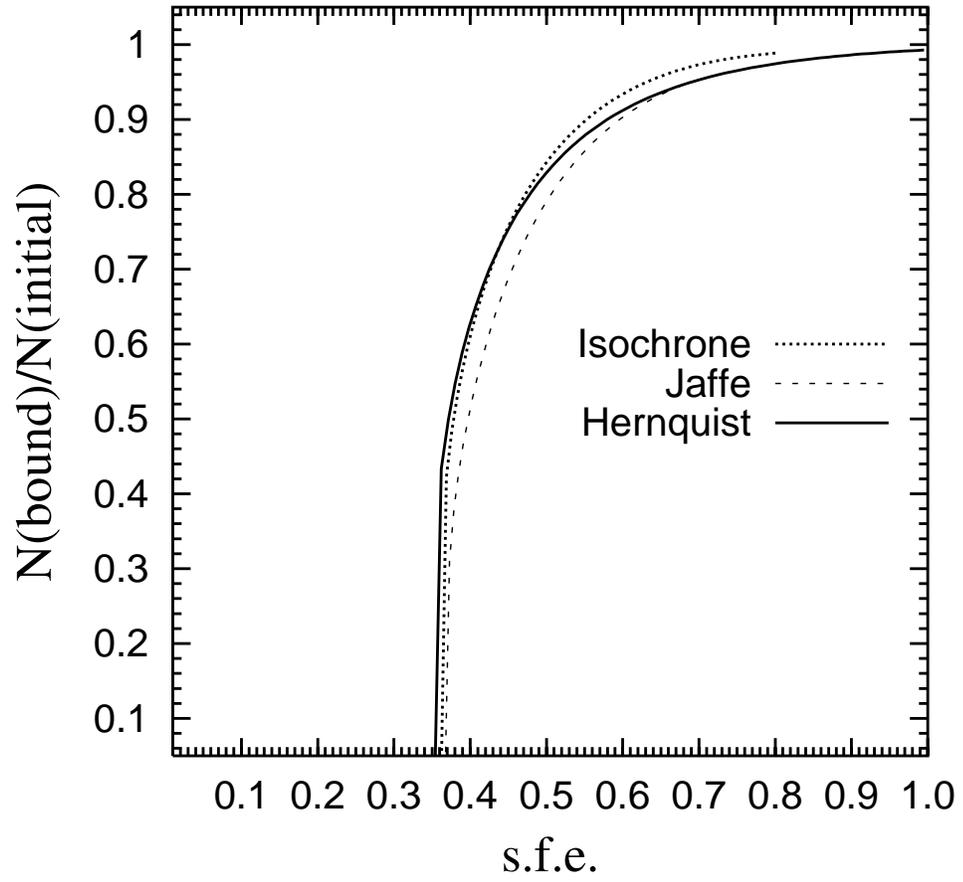}
                  }
\end{picture}
\caption{Bound fraction $M_\star^b/M_\star$ versus \sfe\ for three cuspy models (no harmonic cores) identified in the key.  
All three curves break sharply below \sfe\ = 0.40.} 
\label{fig:hernquist_all}  
\end{figure*} 

\setlength{\unitlength}{1in} 
\begin{figure*}
\begin{picture}(5,5.)(0,0) 
        \put(-.6,.1){\epsfxsize=0.5\textwidth\epsfysize=0.5\textheight
\epsfbox{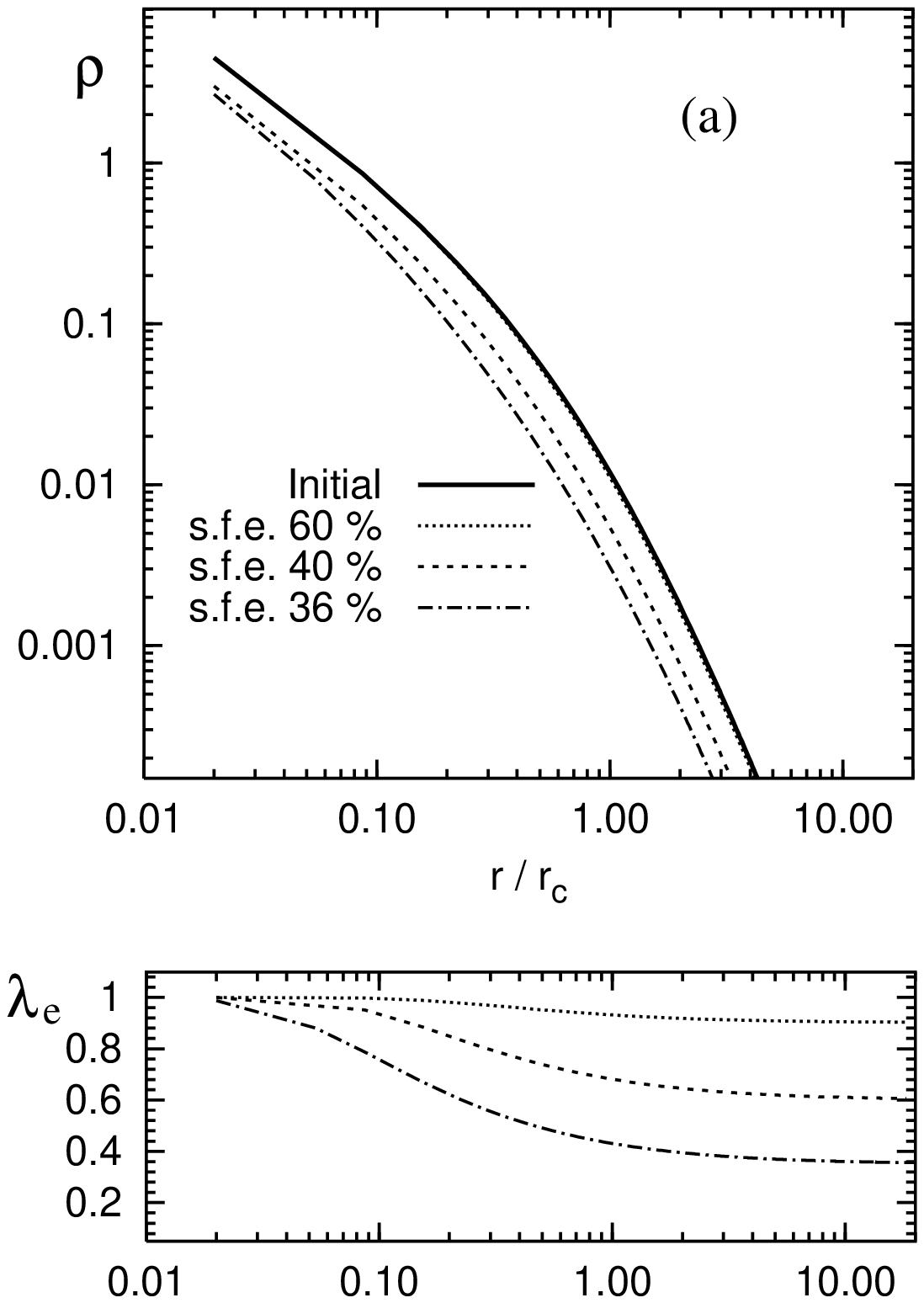} 
                  }
        \put(2.6,0.1){\epsfxsize=0.5\textwidth\epsfysize=0.5\textheight
\epsfbox{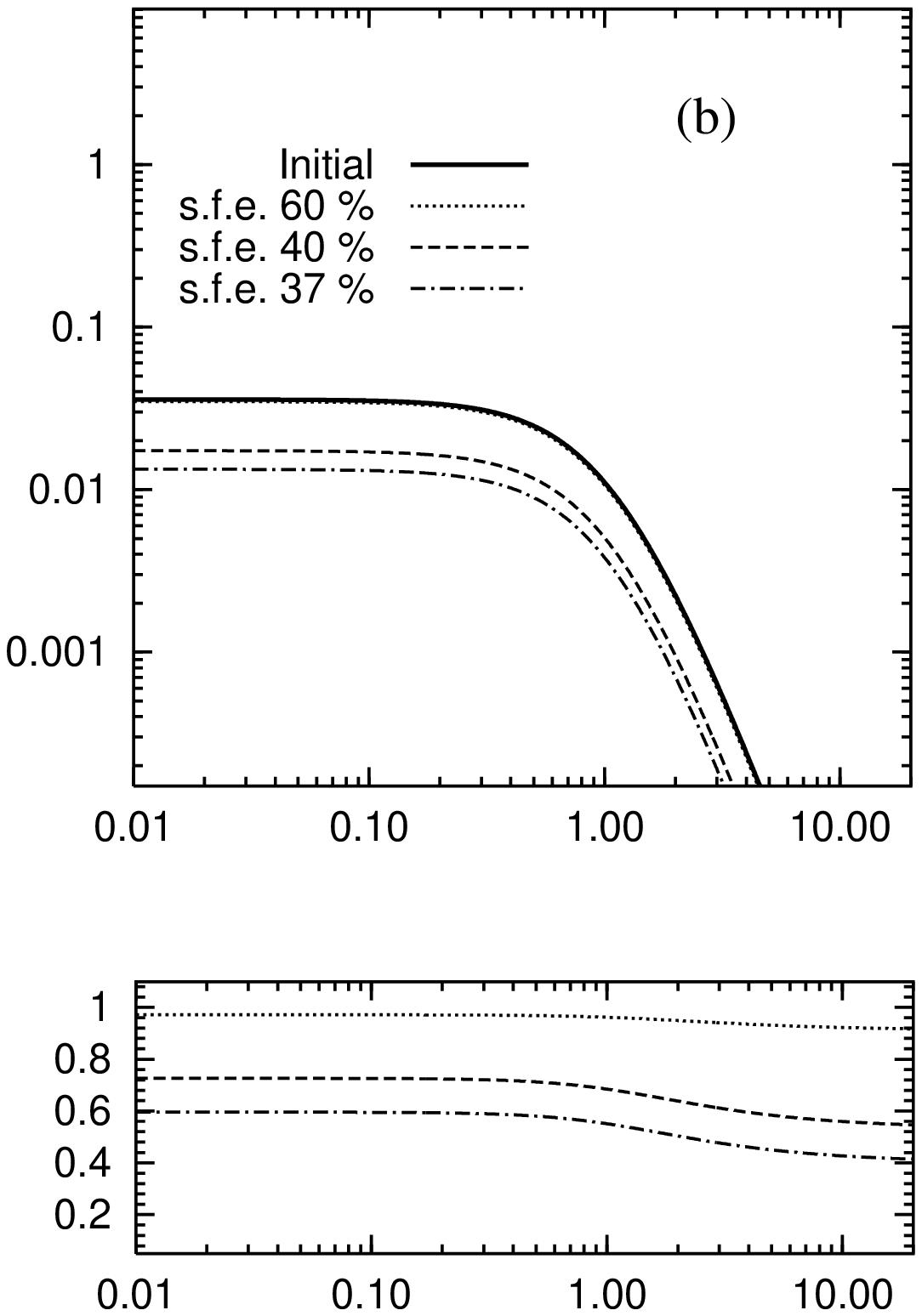} 
                  }
\end{picture}
\caption{Density profiles (top) and fraction of bound stars (bottom) for two models. (a) The Hernquist profile peaks at the centre, $\rho \propto r^{-1}$, whereas on (b) the isochrone model has 
$\rho (r \rightarrow 0) $ = constant. Both have $\rho \propto r^{-4}$ at large distances. When comparing the two models, we find  the bound fraction $\lambda_e$ assumes similar values at large distance for the same \sfe; the bound fraction remains $\approx 1$ at the centre even for low \sfe\ in the case of the Hernquist model, but drops significantly for the isochrone solution.} 
\label{fig:hernquist}  
\end{figure*} 

\end{document}

%% file: mymacros.tex
\newcommand{\getinptfile}
{\typein[\inptfile]{Input file name}
\input{\inptfile}}

\newcommand{\mysummary}[2]{\noi {\bf SUMMARY}#1 \\ \noi \sl #2 \\ \capline 
	\hspace{-.13in} \raisebox{.0in}{$\sqcap$} \rm }  
\newcommand{\mycaption}[2]{\caption[#1]{\footnotesize #2}} 
\newcommand{\capline}{\mbox{}\hrulefill}
\newcommand{\mysection}[2]{ 
\section{\uppercase{\normalsize{\bf #1}}} \def\junksec{{#2}} } %
\newcommand{\mychapter}[2]{ \chapter{#1} \def\junkchap{{#2}}  
\def\thesection{\arabic{chapter}.\arabic{section}}
\def\thesubsection{\thesection.\arabic{subsection}}
\def\thesubsubsection{\thesubsection.\arabic{subsubsection}}
\def\theequation{\arabic{chapter}.\arabic{equation}}
\def\thefigure{\arabic{chapter}.\arabic{figure}}
\def\thetable{\arabic{chapter}.\arabic{table}}
}
\newcommand{\mysubsection}[2]{ \subsection{#1} \def\junksubsec{{#2}} }
\def\thenote{\addtocounter{footnote}{1}$^{\scriptstyle{\arabic{footnote}}}$ }

\newcommand{\myfm}[1]{\mbox{$#1$}}
\def\spose#1{\hbox to 0pt{#1\hss}}	
\def\ltabout{\mathrel{\spose{\lower 3pt\hbox{$\mathchar"218$}} 
     \raise 2.0pt\hbox{$\mathchar"13C$}}}
\def\gtabout{\mathrel{\spose{\lower 3pt\hbox{$\mathchar"218$}}
     \raise 2.0pt\hbox{$\mathchar"13E$}}}
\newcommand{\ltsim}{\raisebox{-0.5ex}{$\;\stackrel{<}{\scriptstyle \backslash}\;$}}
\newcommand{\simlt}{\ltsim}
\newcommand{\simgt}{\gtsim}
%
\newcommand{\unit}[1]{\ifmmode \:\mbox{\rm #1}\else \mbox{#1}\fi}
\newcommand{\ze}{\ifmmode \mbox{z=0}\else \mbox{$z=0$ }\fi }

%
\newcommand{\boldv}[1]{\ifmmode \mbox{\boldmath $ #1$} \else 
 \mbox{\boldmath $#1$} \fi}
%
\renewcommand{\sb}[1]{_{\rm #1}}%
\newcommand{\expec}[1]{\myfm{\left\langle #1 \right\rangle}}
\newcommand{\mone}{\myfm{^{-1}}}
\newcommand{\half}{\myfm{\frac{1}{2}}}
\newcommand{\nth}[1]{\myfm{#1^{\small th}}}
\newcommand{\ten}[1]{\myfm{\times 10^{#1}}}
\newcommand{\abs}[1]{\mid\!\! #1 \!\!\mid}
\newcommand{\as}{a_{\ast}}
\newcommand{\asr}{(a_{\ast}^{2}-R_{\ast}^{2})}
\newcommand{\bvm}{\bv{m}}
\newcommand{\calf}{{\cal F}}
\newcommand{\calI}{{\cal I}}
\newcommand{\calm}{{v/c}}
\newcommand{\calminf}{{(v/c)_{\infty}}}
\newcommand{\calQ}{{\cal Q}}
\newcommand{\calR}{{\cal R}}
\newcommand{\calw}{{\it W}}
\newcommand{\co}{c_{o}}
\newcommand{\cs}{C_{\sigma}}
\newcommand{\cst}{\tilde{C}_{\sigma}}
\newcommand{\cv}{C_{v}}
\def\dbar{{\mathchar '26\mkern-9mud}}	
\newcommand{\deldelr}{\frac{\partial}{\partial r}}
\newcommand{\deldelR}{\frac{\partial}{\partial R}}
\newcommand{\deldeltheta}{\frac{\partial}{\partial \theta} }
\newcommand{\deldelphi}{\frac{\partial}{\partial \phi} }
\newcommand{\ddotrc}{\ddot{R}_{c}}
\newcommand{\ddotxc}{\ddot{x}_{c}}
\newcommand{\dotrc}{\dot{R}_{c}}
\newcommand{\dotxc}{\dot{x}_{c}}
\newcommand{\Estar}{E_{\ast}}
\newcommand{\grpsi}{\Psi_{\ast}^{\prime}}
\newcommand{\kboltz}{k_{\beta}}
\newcommand{\levi}[1]{\epsilon_{#1}}
\newcommand{\limaso}[1]{$#1 ( a_{\ast}\rightarrow 0)\ $}
\newcommand{\limasinfty}[1]{$#1 ( a_{\ast}\rightarrow \infty)\ $}
\newcommand{\limrinfty}[1]{$#1 ( R\rightarrow \infty,t)\ $}
\newcommand{\limro}[1]{$#1 ( R\rightarrow 0,t)\ $}
\newcommand{\limrso}[1]{$#1 (R_{\ast}\rightarrow 0)\ $}
\newcommand{\limxo}[1]{$#1 ( x\rightarrow 0,t)\ $}
\newcommand{\limxso}[1]{$#1 (\xs\rightarrow 0)\ $}
\newcommand{\ls}{l_{\ast}}
\newcommand{\Ls}{L_{\ast}}
\newcommand{\mean}[1]{<#1>}
\newcommand{\ms}{m_{\ast}}
\newcommand{\Ms}{M_{\ast}}
\def\nb{{\sl N}-body }
\def\nbt{{\sf NBODY2} }
\def\nb1{{\sf NBODY1} }
\newcommand{\nuoned}{\nu\sb{1d}}
\newcommand{\ra}{\rightarrow}
\newcommand{\Ra}{\Rightarrow}
\newcommand{\rc}{r_{c} } 
\newcommand{\Rc}{R_{c} } 
\newcommand{\res}[1]{{\rm O}(#1)}
\newcommand{\rnsa}{(r^{2}-a^{2})}
\newcommand{\Rnsa}{(R^{2}-a^{2})}
\newcommand{\rs}{r_{\ast}}
\newcommand{\Rs}{R_{\ast}}
\newcommand{\Rsa}{(R_{\ast}^{2}-a_{\ast}^{2})}
\newcommand{\sa}{\sigma } 
\newcommand{\sac}{\sigma_{c} } 
\newcommand{\sas}{\sigma_{\ast} } 
\newcommand{\sasp}{\sigma^{\prime}_{\ast}}
\newcommand{\saxs}{\sigma_{\ast} } 
\newcommand{\sech}{{\rm sech}}
\newcommand{\tff}{t\sb{ff}} 
\newcommand{\ti}{\tilde}
\newcommand{\trel}{t\sb{rel}}
\newcommand{\ts}{\tilde{\sigma} } 
\newcommand{\tss}{\tilde{\sigma}_{\ast} } 
\newcommand{\vcol}{v\sb{col}}
\newcommand{\vs}{v_{\ast}  } 
\newcommand{\vsp}{v^{\prime}_{\ast}}
\newcommand{\vxs}{v_{\ast}  } 
\newcommand{\xs}{x_{\ast}}
\newcommand{\xc}{x_{c} } 
\newcommand{\xistar}{\xi_{\ast}}
\newcommand{\rmd}{\ifmmode \:\mbox{{\rm d}}\else \mbox{ d}\fi }
\newcommand{\rmD}{\ifmmode \:\mbox{{\rm D}}\else \mbox{ D}\fi }
\newcommand{\valfven}{v_{{\rm Alfv\acute{e}n}}}

%
\newcommand{\noi}{\noindent}
\newcommand{\bc}{boundary condition }
\newcommand{\bcs}{boundary conditions }
\newcommand{\Bcs}{Boundary conditions }
\newcommand{\lhs}{left-hand side }
\newcommand{\rhs}{right-hand side }
\newcommand{\wrt}{with respect to }
\newcommand{\iras}{{\sl IRAS }}
\newcommand{\cobe}{{\sl COBE }}
\newcommand{\Oh}{\myfm{\Omega h}}
%
\newcommand{\etal}{{\em et al.\/ }}
\newcommand{\eg}{{\em e.g.\/ }}
\newcommand{\etc}{{\em etc.\/ }}
\newcommand{\ie}{{\em i.e.\/ }}
\newcommand{\viz}{{\em viz.\/ }}
\newcommand{\cf}{{\em cf.\/ }}
\newcommand{\via}{{\em via\/ }}
\newcommand{\apriori}{{\em a priori\/ }}
\newcommand{\adhoc}{{\em ad hoc\/ }}
\newcommand{\viceversa}{{\em vice versa\/ }}
\newcommand{\versus}{{\em versus\/ }}
\newcommand{\qed}{{\em q.e.d. \/}}
\newcommand{\<}{\thinspace}
%
\newcommand{\km}{\unit{km}}
\newcommand{\kms}{\unit{km~s\mone}}
\newcommand{\kmsa}{\unit{km~s\mone~arcmin}}
\newcommand{\kpc}{\unit{kpc}}
\newcommand{\mpc}{\unit{Mpc}}
\newcommand{\hkpc}{\myfm{h\mone}\kpc}
\newcommand{\hmpc}{\myfm{h\mone}\mpc}
\newcommand{\parsec}{\unit{pc}}
\newcommand{\cm}{\unit{cm}}
\newcommand{\yr}{\unit{yr}}
\newcommand{\au}{\unit{A.U.}}
\newcommand{\AU}{\au}
\newcommand{\gm}{\unit{g}}
\newcommand{\solar}{\myfm{_\odot}}
\newcommand{\solarm}{\unit{M\olar}}
\newcommand{\Lsun}{\unit{L\solar}}
\newcommand{\Rsun}{\unit{R\solar}}
\newcommand{\seconds}{\unit{s}}
\newcommand{\micro}{\myfm{\mu}}
\newcommand{\micrometer}{\micro\mbox{\rm m}}
\newcommand{\Mdot}{\myfm{\dot M}}
%
%
%
\newcommand{\dgr}{\myfm{^\circ} }
\newcommand{\ddgr}{\mbox{\dgr\hskip-0.3em .}}
\newcommand{\mnt}{\mbox{\myfm{'}\hskip-0.3em .}}
\newcommand{\scnd}{\mbox{\myfm{''}\hskip-0.3em .}}
\newcommand{\hr}{\myfm{^{\rm h}}}
\newcommand{\dhr}{\mbox{\hr\hskip-0.3em .}}
%
%
%
%
%
%
%
\newcommand{\refindent}{\par\noindent\hangindent=0.5in\hangafter=1}
\newcommand{\figpar}{\par\noindent\hangindent=0.7in\hangafter=1}
%
%

\newcommand{\mybiblio}{\vspace{1cm}
		       \setcounter{subsection}{0}
		       \addtocounter{section}{1}
		       \def\junksec{References} 
 }

%
%
%

%
%
%
%
%

\newcommand{\vol}[2]{ {\bf#1}, #2}
\newcommand{\jour}[4]{#1. {\it #2\/}, {\bf#3}, #4}
\newcommand{\physrevd}[3]{\jour{#1}{Phys Rev D}{#2}{#3}}
\newcommand{\physrevlett}[3]{\jour{#1}{Phys Rev Lett}{#2}{#3}}
\newcommand{\aaa}[3]{\jour{#1}{A\&A}{#2}{#3}}
\newcommand{\aaarev}[3]{\jour{#1}{A\&A Review}{#2}{#3}}
\newcommand{\aaas}[3]{\jour{#1}{A\&A Supp.}{#2}{#3}}
\newcommand{\aj}[3]{\jour{#1}{AJ}{#2}{#3}}
\newcommand{\apj}[3]{\jour{#1}{ApJ}{#2}{#3}}
\newcommand{\apjl}[3]{\jour{#1}{ApJ Lett.}{#2}{#3}}
\newcommand{\apjs}[3]{\jour{#1}{ApJ Suppl.}{#2}{#3}}
\newcommand{\araa}[3]{\jour{#1}{ARAA}{#2}{#3}}
\newcommand{\mn}[3]{\jour{#1}{MNRAS}{#2}{#3}}
\newcommand{\mnras}{\mn}
\newcommand{\jgeo}[3]{\jour{#1}{Journal of Geophysical Research}{#2}{#3}}
\newcommand{\qjras}[3]{\jour{#1}{QJRAS}{#2}{#3}}
\newcommand{\nat}[3]{\jour{#1}{Nature}{#2}{#3}}
\newcommand{\pasa}[3]{\jour{#1}{PAS Australia}{#2}{#3}}
\newcommand{\pasj}[3]{\jour{#1}{PAS Japan}{#2}{#3}}
\newcommand{\pasp}[3]{\jour{#1}{PAS Pacific}{#2}{#3}}
\newcommand{\rmp}[3]{\jour{#1}{Rev. Mod. Phys.}{#2}{#3}}
\newcommand{\science}[3]{\jour{#1}{Science}{#2}{#3}}
\newcommand{\vistas}[3]{\jour{#1}{Vistas in Astronomy}{#2}{#3}}

%% file: paper2.bbl
\begin{thebibliography}{}

\bibitem{} Adams F.C., 2000, ApJ, 542, 964

\bibitem{} Baumgardt H., 2001, MNRAS, 325, 1323 

\bibitem{} Binney J.J., Tremaine S.D., 1987, Galaxy Dynamics, Princeton: Pinceton University Press, 733 ff (BT+87)


\bibitem{} Boily C.M., Kroupa P., 2002, MNRAS, in the press (Paper I) 

\bibitem{} Boily C.M., Kroupa P., Pe\~narrubia-Garrido J., 2001, New Ast, 6, 27

\bibitem{} Brown A.G.A., 2001, Rev. Mex. Astronomy, 11, 89  (see also astro-ph/0101207) 
 
\bibitem{} Clarke C.J., Bonnell I.A., Hillenbrand L., 2000, in Protostars and Planets IV, ed. V. Mannings, A.P. Boss \& S.S. Russell, 
Tucson: University of Arizona Press, p. 151

%


\bibitem{} Fellhauer M., Kroupa P., 
 Baumgardt H. et al., 2000, New Ast, 5, 305

\bibitem{} Geyer M.P., Burkert A., 2001, MNRAS, 323, 98

\bibitem{} Goodwin S. P., 1997, MNRAS, 284, 785

\bibitem{} H\'enon M., 1960, Annales d'Astrophysique, 23, 474

\bibitem{} Hernquist L., 1990, ApJ, 356, 359 

\bibitem{} Hernquist L., 1993, APJS, 86, 389

\bibitem{} Hills J., 1980, ApJ, 225, 986 

\bibitem{} Jaffe W.,  1983, MNRAS, 202, 995 

\bibitem{} King I.R., 1966, AJ, 71, 64  

\bibitem{} Kroupa P., Aarseth S.J., Hurley J., 2001, MNRAS, 321, 699

\bibitem{} Lada C.J., Margulis M.,  Deardorn D., 1984, ApJ, 285, 141

\bibitem{} Lada E. A., 1999, in NATO Science Series C Vol. 450, The
Origins of Stars and Planetary Systems, ed. C.J. Lada \& N.D. Kylafis
(Dordrecht: Kluwer), 441

\bibitem{} Meylan G., Heggie D.C., 1997, A \& A Rev, 8, 1

\bibitem{} Michie R., 1963, MNRAS, 125, 127, \S 4 

\bibitem{} Nagata T., Woodward C.E., Shure M., Kobayashi N., 1995, AJ, 109, 1676 

\bibitem{} Plummer H.C., 1911, MNRAS, 71, 460

\bibitem{} Press W.H., Teukolsky S.A.,  Vetterling W.T., Flannery B., 1992, Numerical Recipes: The Art of Scientific 
  Computing, Cambridge: University Press, 963 ff

\bibitem{} Spitzer L., 1987, Dynamical Evolution of Star clusters,
Princeton: Princeton University Press, 232 ff 

\bibitem{} Terlevich E., 1987, MNRAS, 224, 193 

\bibitem{} Testi L. et al, 1997, A\&A, 320, 159 

\bibitem{} Testi L. et al, 1999, A\&A, 342, 514 

\bibitem{} Weinberg M.D., 1993, in ASP Conf. Series Vol. 48, ed. H. Smith \& J.P. Brodie, San Francisco: PASP, 689

\end{thebibliography}
